\documentclass[
jcp,
amsmath,amssymb,
preprint,
]{revtex4-1}

\usepackage{graphicx}
\usepackage{dcolumn} 
\usepackage{bm}      
\usepackage{multirow} 
\usepackage{subfigure}

\begin{document}

\title[Sample title]{Dependence of the fragility of a glass former on the softness of interparticle interactions}
\author{Shiladitya Sengupta$^{1}$, Filipe Vasconcelos$^{2}$,
Fr\'{e}d\'{e}ric Affouard$^{2}$, Srikanth Sastry$^{1}$}
\affiliation{
$^{1}$ Theoretical Sciences Unit, Jawaharlal Nehru Centre for Advanced Scientific Research,
Jakkur Campus, Bangalore 560 064, India.\\
$^{2}$ Unit\'{e} Mat\'{e}riaux et Transformations (UMET), UMR CNRS 8207,
Universit\'{e} Lille Nord de France, Villeneuve d\'{} Ascq, France.
}

\date{\today}

\begin{abstract}
We study the influence of the softness of the interparticle
interactions on the fragility of a glass former, by considering three
model binary mixture glass formers. The interaction potential between
particles is a modified Lennard-Jones type potential, with the
repulsive part of the potential varying with an inverse power $q$ of
the interparticle distance, and the attractive part varying with an
inverse power $p$. We consider the combinations (12,11) (model I),
(12,6) (model II) and (8,5) (model III) for (q,p) such that the
interaction potential becomes softer from model I to III. We evaluate
the kinetic fragilities from the temperature variation of diffusion
coefficients and relaxation times, and a thermodynamic fragility from
the temperature variation of the configuration entropy. We find that
the kinetic fragility increases  with increasing softness of the
potential, consistent with previous results for these model
systems, but at variance with the thermodynamic fragility, which
decreases with increasing softness of the interactions, as well as
expectations from earlier results. We rationalize our results by
considering the full form of the Adam-Gibbs relation, which requires,
in addition to the temperature dependence of the configuration
entropy, knowledge of the high temperature activation energies ino
rder to determine fragility. We show that consideration of the
scaling of the high temperature activation energy with the liquid
density, analyzed in recent studies, provides a partial rationalization of the observed behavior. 
\end{abstract}

\pacs{Valid PACS appear here}

%Journal of Chemical Physics 135, 194503 (2011)
%doi:10.1063/1.3660201

                              
\maketitle

\section{Introduction}

The temperature variation of relaxation times, viscosity and diffusion
coefficient in glass forming liquids upon approaching the glass
transition has been studied for a wide variety of substances. Near the
glass transition, these quantities show a rapid increase, but with a
rate of change that is different  for different substances. The
rapidity of rise of relaxation times near the glass transition has
been quantified by ``fragility'', introduced and analyzed extensively
by Angell \cite{fragility_angell}, which has proved to be useful in
organizing and understanding the diversity of behavior seen in glass
formers. Fragility has been defined in a variety of ways. Two of the
popular definitions are in terms of the ``steepness index'' $m$, and
the fragility defined using Vogel-Fulcher-Tammann (VFT) fits to viscosity
and relaxation time data. 

The steepness index of fragility is defined from the so-called Angell
plot as the slope ($m$) of logarithm of the viscosity ($\eta$) or
relaxation time ($\tau$) at $T=T_{g}$, with respect to the scaled
inverse temperature $T_g/T$ where $T_{g}$ is the laboratory glass
transition temperature:

\begin{equation}
m = \left( \frac{d \log \tau}{d (\frac{T_{g}}{T})}\right )_{T=T_{g}} \label{eqn:kinfr1}
\end{equation} 

We refer to the fragilities defined from transport quantities and relaxation times as {\it kinetic} fragilities, to be distinguished from {\it thermodynamic} fragilities defined later. A kinetic fragility may also be defined from a VFT fit of the relaxation times, 

\begin{equation}
\tau(T) = \tau_{0} \exp\left[\frac{1}{K_{VFT}(\frac{T}{T_{VFT}}-1)} \right] \label{eqn:kinfr2}
\end{equation} 

which defines the kinetic fragility $K_{VFT}$ and the divergence temperature $T_{VFT}$. 

Despite considerable research effort
\cite{fragility_angell,speedy,pap:AG-Sastry,wales,tarjus,ruocco,sokolov,pap:BordatPRL,pap:BordatJNCS, pap:Dudowicz-JPCB-2005, pap:Dudowicz-JCP-2005, douglas, francis, sneha,tanaka, pap:Mattsson-etal, pap:Angell-news-views},
and the observation of many empirical correlations between fragility
and other material properties, a fully satisfactory understanding of
fragility hasn't yet been reached. Such understanding has been sought,
broadly, along two lines. The first is a conceptual understanding of
fundamental quantities that may govern fragility. An example of this
kind is the use of the potential energy landscape approach in
combination with Adam Gibbs (AG) relation \cite{AdamGibbs} between relaxation time and
configuration entropy [Eq. \ref{eqn:AG}] to relate features of the
energy landscape of a glass former to the fragility. The Adam-Gibbs
relation
\begin{equation}
\tau(T) = \tau_{0} \exp (\frac{\delta \mu S^{*} k_{B}^{-1}}{T S_c} )
\label{eqn:AG}
\end{equation} 
relates the temperature dependence of the relaxation times to the
temperature change in the configuration entropy $S_c$, where $\delta \mu$ is an activation free energy for particle rearrangements, and $S^{*}$ is the configurational entropy of cooperatively rearranging regions invoked in Adam-Gibbs theory. If $A \equiv 
\delta \mu S^{*} k_{B}^{-1}$ has no significant role to play in
determining the fragility of a substance, it is the temperature variation 
of $T S_c$ that dictates the fragility. If the T-dependence of $S_c$ is given by 
\begin{equation}
TS_{c} = K_{T}\left(\frac{T}{T_{K}} - 1\right), \label{eqn:pelfrag}
\end{equation}
the Adam-Gibbs relation yields the VFT relation, with the identification 
$K_{VFT} = K_T/A$, $T_{VFT} = T_{K}$. Thus, $K_T$ is a thermodynamic index of fragility. 

In what follows, we use Eq.s \ref{eqn:kinfr2} and \ref{eqn:AG} which
describe our simulation data well, as we demonstrate. However, our
discussion does not depend crucially on the strict validity of the VFT
temperature dependence near the glass transition, or the divergence of
relaxation times at finite temperature; both these features have been
questioned by various investigations and alternative forms to the VFT
temperature dependence have been proposed \cite{Rossler,Chandler,Dyre}.

In potential energy landscape approach \cite{pap:PEL-Sciortino,pap:PEL-Heuer} 
configuration entropy is associated with the number of local potential energy minima or {\it inherent structures} (IS) \cite{inh}, and can be computed in terms of
parameters describing the energy landscape \cite{pap:AG-Sastry}.
Hence thermodynamic fragility can be understood in terms of parameters
of the potential energy landscape, namely the distribution of inherent
structures and the dependence of
the vibrational or basin entropy corresponding to inherent structures on theie energies. Although the
exact temperature dependence of the configuration entropy depends on
detailed properties of the distribution of inherent structures, and
$K_T$ is not a constant even in the simplest case, such analysis does
yield insight into the relationship between the energy landscape
features and fragility. To a first approximation, the broader the
distribution of energies of inherent
  structures, the larger the fragility of a glass former
\cite{pap:AG-Sastry}. Going beyond such analysis, one needs to also
understand the behavior of the prefactor $A$, which is related to the
high temperature activation energy
\cite{tarjus,schroder,douglas,ruocco}. To the extent that the
Adam-Gibbs relation quantitatively describes the temperature
dependence of the relaxation times, such analysis provides a route to
a fundamental understanding of fragility in terms of the phase space
properties of a substance. 

However, such a conceptual understanding does not directly address the
dependence of fragility on specific, controllable material properties,
an understanding that is desirable from the perspective, {\it e. g.},
of materials design. The investigation of the dependence of fragility
on the nature of molecular architecture and intermolecular interactions
defines therefore a second distinct line of investigation, which has
been pursued by various groups. For example, Dudowicz, Freed and Douglas
\cite{pap:Dudowicz-JPCB-2005, pap:Dudowicz-JCP-2005}
have investigated the role 
of backbone and side group stiffness in determining the fragility of polymer glass
formers. In another recent example, from an experimental investigation
on deformable colloidal suspensions, Mattsson {\it et al}
\cite{pap:Mattsson-etal,pap:Angell-news-views} suggested that increasing the softness of the
colloidal particles should decrease the fragility of the colloidal
suspensions, and that such a principle should be more generally
applicable. Indeed, this conclusion is consistent with that of Douglas
and co-workers \cite{pap:Dudowicz-JPCB-2005} that the ability to better pack
molecules leads to lower fragilities. In energy landscape terms, one
may understand this conclusion as implying that molecules that pack
well together will have narrower distributions of inherent structure
energies. 

\begin{figure}[h!]
\begin{center}
\vspace{4mm}
\includegraphics[width=6cm, height=5cm]{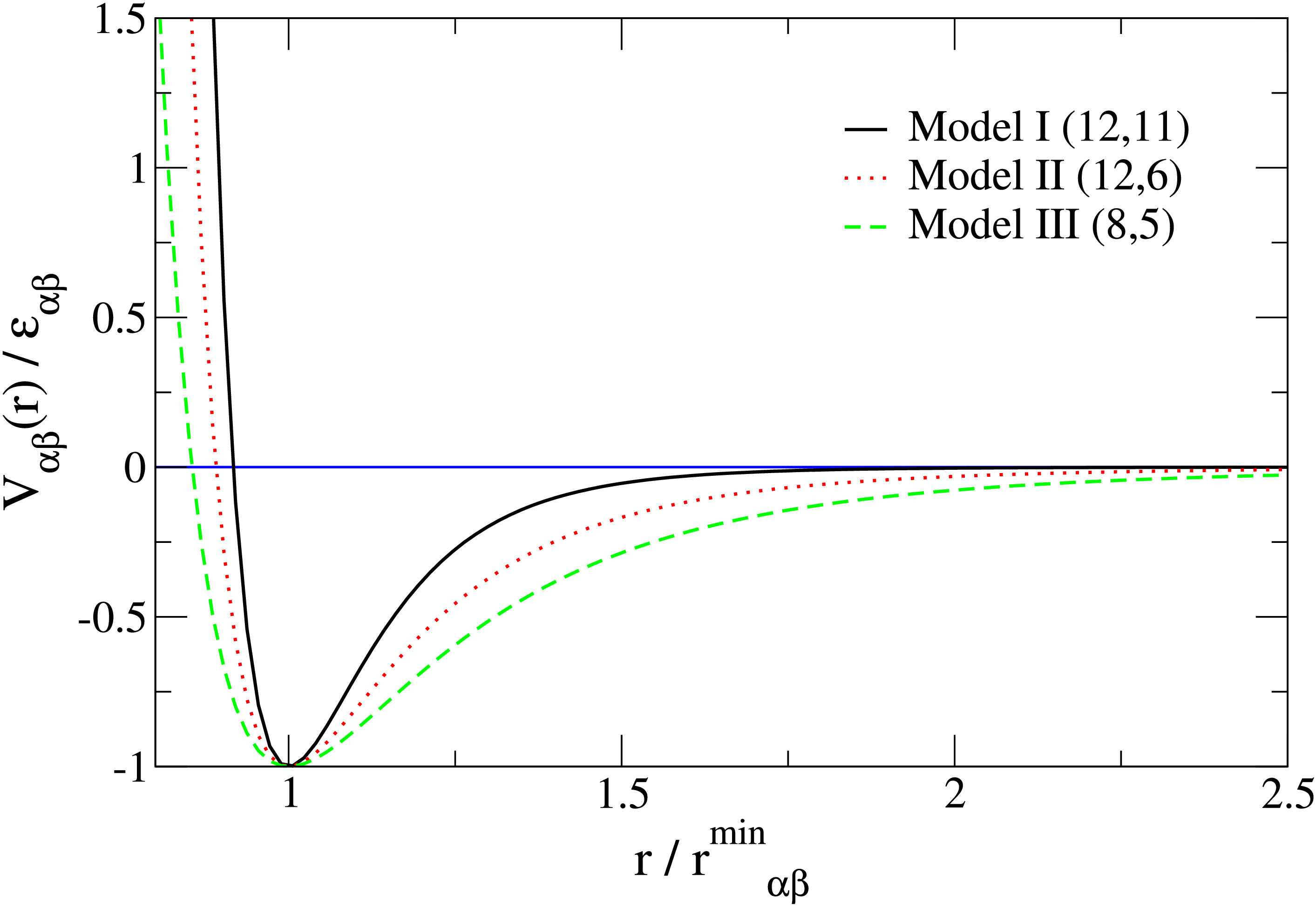}
\caption{Comparison of interaction potential $V_{\alpha\beta}$ without
  truncation for the three different potentials used in the present
  study. $r^{min}_{\alpha\beta}$ are the positions of the minima of
  the interaction potentials.}\label{fig:comp-pot}
\end{center}
\end{figure}

\begin{figure}[h!]
\begin{center}
\includegraphics[width=6cm, height=5cm]{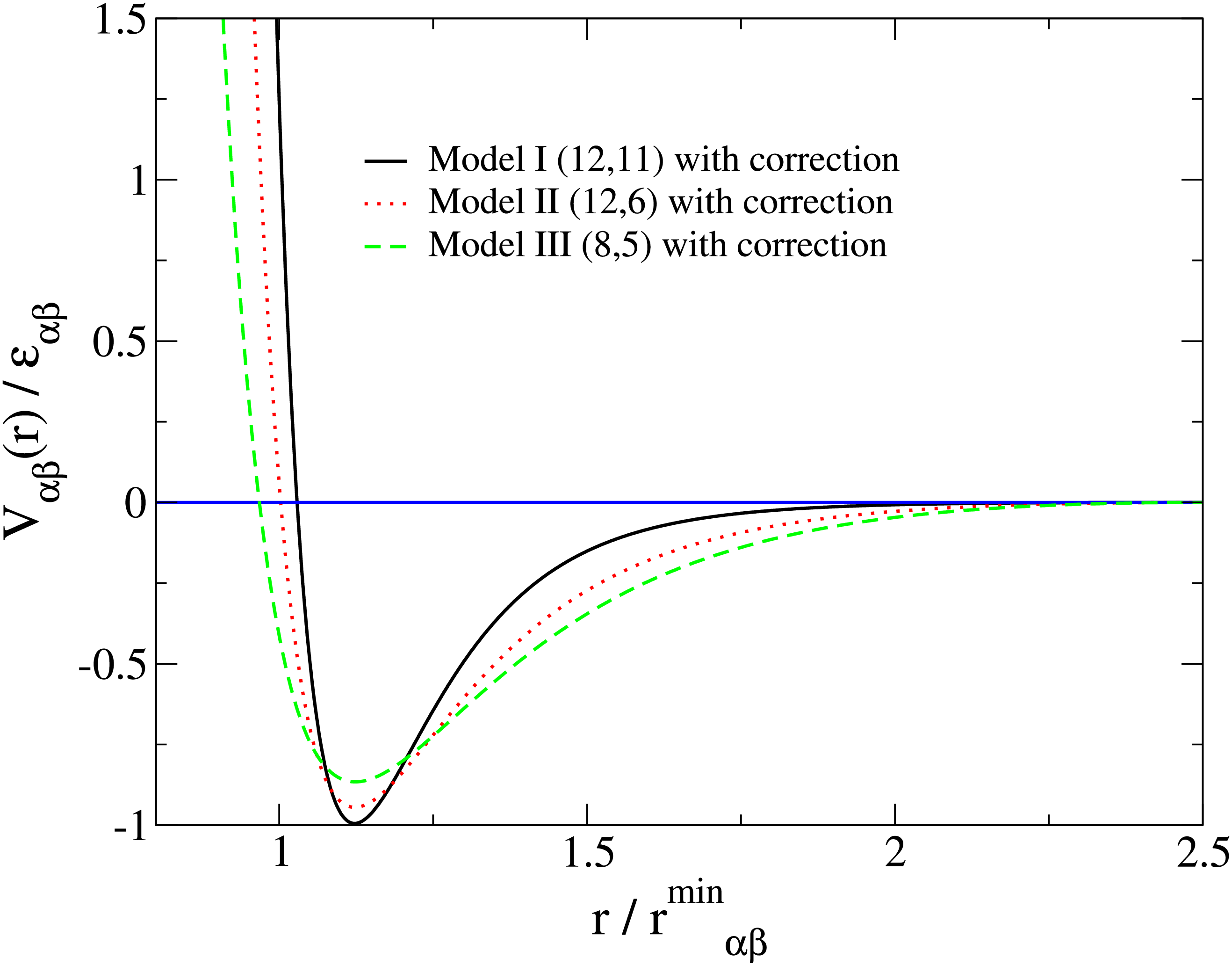}
\caption{Comparison of the interaction potential $V_{\alpha\beta}$ with truncation for the three different potentials. 
$r^{min}_{\alpha\beta}$ are the positions of the minima of the interaction potentials. }\label{fig:comp-pot-correction} 
\end{center}
\end{figure}

The influence of the softness of interaction on the fragility was also 
investigated some time ago {\it via} computer simulations of model glass formers 
by Bordat {\it et al} ~\cite{pap:BordatPRL, pap:BordatJNCS}. They considered 
a binary mixture of particles interacting {\it via}
 generalized Lennard Jones potentials, of the form 
\begin{equation} 
V(r) = {\epsilon \over (q - p)} \left( p ({\sigma \over r})^q - q ({\sigma \over r})^p\right)
\label{eqn:potential}
\end{equation}
for combinations of the exponents $(q,p)$ of repulsive and attractive
parts of the potential (12,11), (12,6) and (8,5). These combinations,
corresponding to models labeled I, II and III, have decreasing
curvatures at the minimum of the potential, and thus increasing
softness.  By evaluating the {\it kinetic} fragility of these models
(the steepness index defined above), Bordat {\it et al} found that
\emph{increasing} softness of the interaction potential
\emph{increases} the kinetic fragility \cite{pap:BordatPRL,pap:BordatJNCS}. 

The trend found by Bordat {\it et al} therefore is apparently not
consistent with expectations arising from the other studies mentioned,
although the nature of the changes in the interactions considered are
not strictly the same. In order to understand better the relationship
between the nature of the intermolecular interactions and fragility,
in the present work we calculate the kinetic fragility $K_{VFT}$ using
computer simulation data of the diffusion coefficient, and relaxation
times obtained by a number of different means. We also calculate,
using the procedure in
\cite{pap:AG-Sastry,pap:Sc-Sastry,pap:Sc-Sastry-JPCM}, the
configuration entropy, from which we calculate a thermodynamic
fragility ($K_{T}$). We find that these two fragilities show opposite
trends, with the kinetic fragility increasing with softness, and the
thermodynamic fragility decreasing with softness. In order to
understand this apparent disagreement, we must consider the full form
of the Adam-Gibbs relation, including terms that relate to the high
temperature activation energy. We present our analysis along these
lines below. We focus our analysis here on the role of a specific
feature of the interaction potential, namely the softness, for reasons
stated above. However, fragility in principle depends on a number of
parameters that describe a glass former, which may include pressure,
density {\it etc.}. While our study implicitly includes those factors
that are affected by a change in the interaction potential (keeping
other parameters fixed), we do not attempt here a comprehensive
analysis of all factors that may influence the fragility of a glass
former.
Related questions concerning the change in structure, dynamics and thermodynamics in a glass forming liquid
upon tuning the interaction potential have been addressed in \cite{pap:Shi-etal}

The paper is organized as follows: In Section II we summarize the
computer simulation details. In Section III we describe the methods
used for evaluating the various quantities of interest. In Section IV
we present our results and a discussion of the
results, and Section V contains our conclusions. 

\section{Simulation Details}

\begin{figure}[h!]
\begin{center}
\includegraphics[width=7cm, height=6cm] {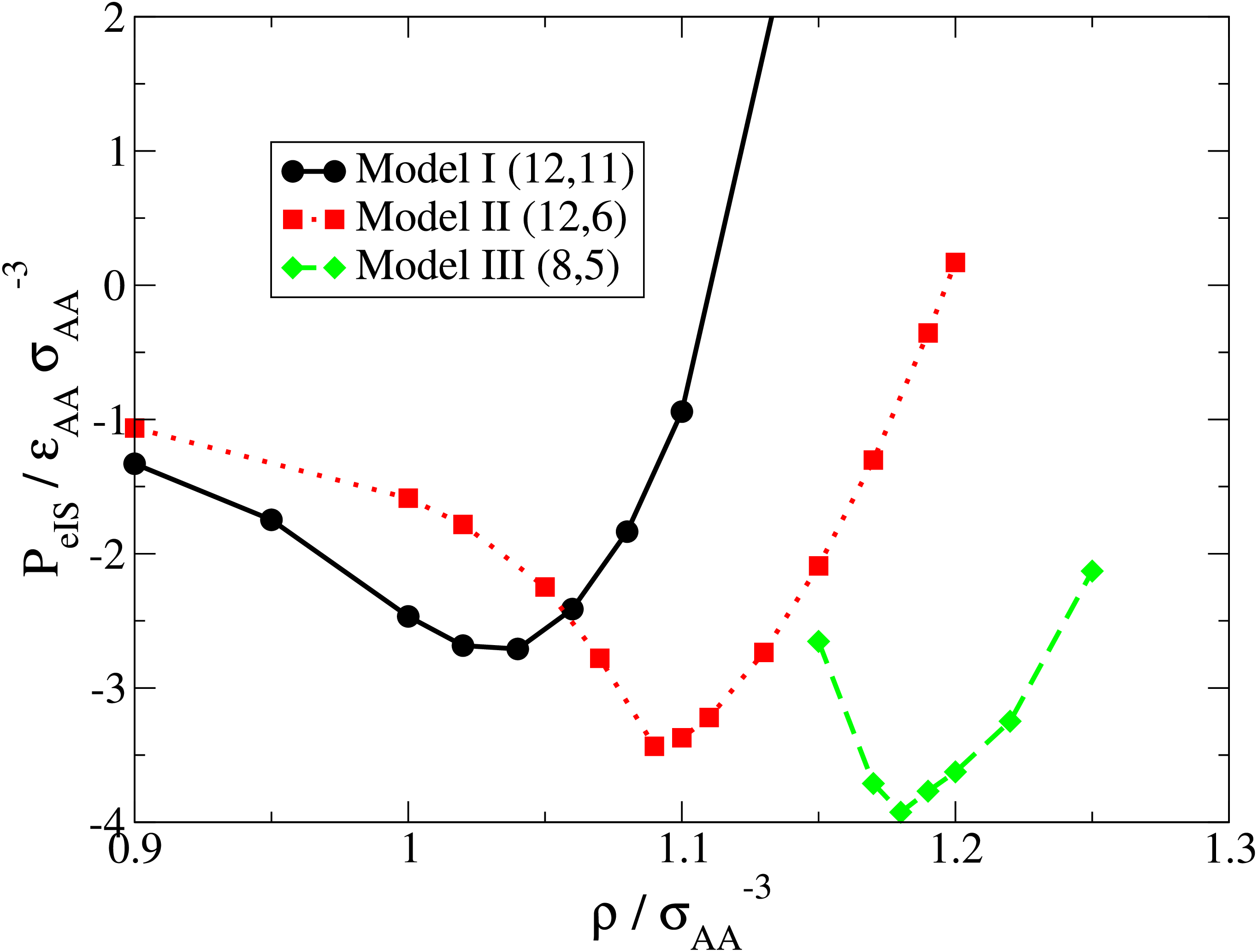}
\caption{Pressure {\it vs.} density for inherent structures (IS). The density minimum for IS pressure occurs at $\rho=1.04,1.09,1.18$ respectively for models I $(12,11)$, II $(12,6)$, and III $(8,5)$. This density defines the lower bound for simulations of the system in the homogeneous liquid state.}\label{p-eIS-vs-T-isotherm-allmodel}
\end{center}
\end{figure}

We have studied a 80:20 binary mixture of modified Lennard Jones
particles in three dimensions. The interaction potential is of the
form given above in Eq. \ref{eqn:potential} with a truncation that  
makes both the potential and force go to zero smoothly at a cutoff distance $r_{c}$. The potential with the truncation is given by 
\begin{eqnarray}
V_{\alpha\beta}(r) &=& \frac{\epsilon_{\alpha\beta}}{q-p}\left[ p( \frac{r_{\alpha\beta}^{min}}{r})^{q} - q( \frac{r_{\alpha\beta}^{min}}{r})^{p}\right]\nonumber\\
&& + c_{1\alpha\beta} r^{2} + c_{2\alpha\beta}, r < r_{c\alpha\beta} \nonumber\\
                 &=& 0, \hspace{5mm}\mbox{otherwise}
\label{eqn:def-pot-corrected}
\end{eqnarray} 
where $\alpha,\beta \in \{A,B\}$.  $r_{\alpha\beta}^{min} = 2^{\frac{1}{6}} \sigma_{\alpha\beta}$  
and $\epsilon_{\alpha\beta}$ are respectively the position and the value of the minimum 
of the pair potential. The correction terms $c_{1\alpha\beta}, c_{2\alpha\beta}$ are determined from the conditions :
\begin{eqnarray}
V_{\alpha\beta}(r_{c\alpha\beta}) &=& 0 \nonumber\\
\left(\frac{d V_{\alpha\beta}}{d r}\right)_{r_{c\alpha\beta}} &=& 0
\end{eqnarray}
The energy and size parameters $\epsilon_{\alpha \beta}$ and $\sigma_{\alpha \beta}$
correspond to those of the Kob-Andersen binary Lennard-Jones model \cite{kob}.
Units of length, energy and time scales are $\sigma_{AA},\epsilon_{AA}$ and
$\sqrt{\frac{\sigma_{AA}^{2}m_{AA}}{\epsilon_{AA}}} $ respectively.
In this unit, $\epsilon_{AB}=1.5$, $\epsilon_{BB}=0.5$,
$\sigma_{AB}=0.80$, $\sigma_{BB}=0.88$. 
The interaction potential was cutoff at $2.5 \sigma_{\alpha\beta}$, 
The three different models  $(12,11)$, $(12,6)$ and $(8,5)$
are shown without and with cutoff in Fig.s \ref{fig:comp-pot} and \ref{fig:comp-pot-correction}.  Molecular
dynamics (MD) simulations were done in a cubic box with periodic
boundary conditions in the constant nummber, volume and temperature  (NVT) ensemble. The integration
time step was in the range $dt = 0.001-0.005$. Temperatures were kept
constant using an algorithm due to Brown and Clarke
\cite{pap:BC}. Simulations were done in the temperature range $T \in
\left[ 0.85, 5\right]$ for $(12,11)$; $T \in \left[ 0.45, 5\right]$
for $(12,6)$ and $T \in \left[ 0.23, 5\right]$ for $(8,5)$ model
respectively. System size were $N=1500, N_{A}=1200$ ($N =$ total
number of particles, $N_{A}=$ number of particles of species $A$) and
the number density was $\rho = 1.2$ (\cite{pap:BordatPRL}, see also Fig. \ref{p-eIS-vs-T-isotherm-allmodel}). 
For all models, one sample per state point above the onset temperature (described below) and three to five samples per state points below the onset temperature were used with runlengths $>100 \tau_{\alpha}$ ($\tau_{\alpha}$ is the relaxation time, described below).

\section{Methods}

In this section, we describe the various quantities that have been calculated and the methods employed for such calculations. 

\subsection{The $\alpha$ relaxation time}

The following measures have been used to extract $\alpha$ relaxation times:\\
\begin{enumerate}
\item

Diffusion coefficient ($D_{A}$) from the mean squared displacement (MSD) of the $A$ type particles.

\item
Relaxation times obtained from the decay of overlap function $q(t)$
using the definition $q(t=\tau_{\alpha},T)/N = 1/e$. The overlap function is a
two-point time correlation function of local density 
\cite{pap:4pt-CD, pap:Ovlap-Glotzer-etal, pap:Ovlap-Donati-etal, pap:Lacevic, pap:Karmakar-PNAS} 
which has been used in many recent studies of slow relaxation, and is defined as: 
\begin{eqnarray}
<q(t)>  &\equiv&  <\int d\vec{r}\rho(\vec{r},t_{0})\rho(\vec{r},t+t_{0})>\nonumber\\
&=&  <\sum_{i=1}^{N} \sum_{j=1}^{N} \delta (\vec{r}_{j}(t_{0}) - \vec{r}_{i}(t+t_{0}))>
\label{eqn:qtdef}
\end{eqnarray}
Here the averaging over time origins $t_{0}$ is implied. The overlap function naturally 
separates into ``self'' and ``distinct'' terms:
\begin{eqnarray}
<q(t)>&=&  <\sum_{i=1}^{N} \delta (\vec{r}_{i}(t_{0}) - \vec{r}_{i}(t+t_{0}))> \nonumber\\
&& +  <\sum_{i} \sum_{j\neq i} \delta (\vec{r}_{i}(t_{0}) - \vec{r}_{j}(t+t_{0}))>\nonumber
\end{eqnarray}
In our work, we consider only the self part of the total overlap function ({\it i.e.} neglect the $i\ne j$ terms in the double summation), based on the observation \cite{pap:Lacevic} that the results obtained from the self part are not significantly different from those obtained by considering the collective overlap function. Thus we use 

\begin{eqnarray}
<q(t)>&\approx&  <\sum_{i=1}^{N} \delta (\vec{r}_{i}(t_{0}) - \vec{r}_{i}(t+t_{0}))>\nonumber
\end{eqnarray}
Further, for numerical computation, the $\delta$ function is approximated by a window function $w(x)$ which defines the condition of ``overlap'' between two particle positions separated by a time interval $t$: 
\begin{eqnarray}
<q(t)> &\approx&  <\sum_{i=1}^{N} w(|\vec{r}_{i}(t_{0}) - \vec{r}_{i}(t_{0}+t)|)>\nonumber\\
w(x) &=& 1, x \leq a \textnormal{  implying ``overlap''}\nonumber\\
     &=& 0 \mbox{   otherwise   }\nonumber
\end{eqnarray}
The time dependent overlap function thus depends on the choice of the cutoff parameter $a$, which we choose to be $0.3$.  This parameter is chosen such that particle positions separated due to small amplitude vibrational motion are treated as the same, or that $a^{2}$ is comparable to the value of the MSD in the plateau between the ballistic and diffusive regimes. 

\item
We have also studied the ``susceptibility'' $\chi_{4}(t)$, defined in terms of the fluctuations in the overlap function as 
\begin{equation}
\chi_{4}(t) = {1 \over N} \left( \langle q(t)^{2} \rangle - \langle q(t) \rangle ^{2} \right)\label{eqn:chi4}
\end{equation}
This quantity can be written as an integral of a higher order, \emph{four point} correlation function  $g_{4}(\vec{r},t)$ \cite{pap:4pt-CD, pap:Ovlap-Glotzer-etal, pap:Ovlap-Donati-etal} widely studied in the context of dynamical heterogeneity:
\begin{eqnarray}
g_{4} (\vec{r},t) &=& \langle \rho(0,0) \rho(\vec{r},0) \rho(0,t) \rho(\vec{r},t) \rangle - \nonumber\\
&& \langle \rho(0,0) \rho(\vec{r},0) \rangle \langle \rho(0,t) \rho(\vec{r},t) \rangle \nonumber\\
\chi_{4}(t) &=& \int d\vec{r} g_{4}(\vec{r},t)
\end{eqnarray} 
The characteristic time $\tau_{4}(T)$ at which the fluctuation ($\chi_{4}(t)$)  is maximum is taken as a measure of relaxation time. 

\item
Relaxation times obtained from the decay of the self intermediate scattering function $F_{s}(k,t)$ using the definition $F_{s}(k,t=\tau_{\alpha},T)=1/e$ at $k \simeq \frac{2 \pi}{r_{min}}$. The self intermediate scattering function is calculated from the simulated trajectory as: 
\begin{equation}
F_{s}(k,t) =\frac{1}{N}\langle \sum_{i=1}^{N} \exp\left(-\imath \vec{k}\cdot (\vec{r}_{i}(t) - \vec{r}_{i}(0))\right)\rangle 
\label{eqn:Fskt}
\end{equation} 
\end{enumerate}

Since the relaxation times from $q(t)$, $\chi_4(t)$ and $F_s(k,t)$ behave very similarly, we discuss further only the time scale obatined from $q(t)$.

\subsection{Characteristic temperature scales}
Dynamics of fragile glass forming liquids show characteristic cross-over from high temperature Arrhenius behaviour to low temperature super Arrhenius behaviour at some characteristic temperature. At this temperature, systems also show cross-over from a “landscape independent” high temperature regime to a “landscape influenced” low temperature regime
\cite{pap:sastry-deb-stil,pap:Sastry-pcc}. We denote this temperature as onset temperature $T_{onset}$. We report the estimates from inherent structure energies in Table \ref{tab:TempScale}.
As the temperature is further lowered, mode coupling theory predicts divergence of relaxation time  $\tau$ as  
$\tau(T) \sim (T-T_{c})^{-\gamma}$ which defines the mode coupling divergence temperature $T_{c}$ which we estimate from both relaxation time and diffusion coefficient (in the form $(D_{A}/T)^{-1}$) \cite{kob}. 
Similarly relaxation times apparently diverge at a second characteristic temperature which we estimate from VFT fits and denote as $T_{VFT}$. Further, configuration entropy becomes zero on extrapolation at a characteristic temperature  (Eq. \ref{eqn:pelfrag}) known as  Kauzmann temperature ($T_{K}$). The AG relation (Eq. \ref{eqn:AG}) predicts that these two temperatures ($T_{VFT}$ and $T_{K}$) to be same. Although we use functional forms that have a temperature of vanishing $S_c$ and diverging relaxation times, these are employed as useful descriptions of the data, without any implied assertion of the expected behavior at temperatures lower than the ones we study. The 
 The values of different characteristic temperatures  for different potentials are tabulated in Table  \ref{tab:TempScale}.

\begin{table}[!]
\caption{\label{tab:TempScale}  Characteristic Temperatures}
\begin{ruledtabular}
\begin{tabular}{cccc}
Quantity & (12,11) & (12,6) & (8,5)\\
\hline
$T_{onset}$  & 1.27 & 0.9 & 0.42 \\
$T_{c}$  from $(\frac{D_{A}}{T})^{-1}$ & 0.77 & 0.42 & 0.22\\  
$T_{c}$  from $q(t)$ & 0.77 & 0.42 & 0.23 \\
$T_{VFT}$ from $(\frac{D_{A}}{T})^{-1}$ & 0.59 & 0.32 & 0.17\\  
$T_{VFT}$ from $q(t)$ & 0.55 & 0.29 & 0.16\\  
$T_{K}$    & 0.54 & 0.28 & 0.16 \\
\end{tabular}
\end{ruledtabular}
\end{table}

\begin{figure}[!]
\begin{center}
\includegraphics[width=8cm,height=6cm]{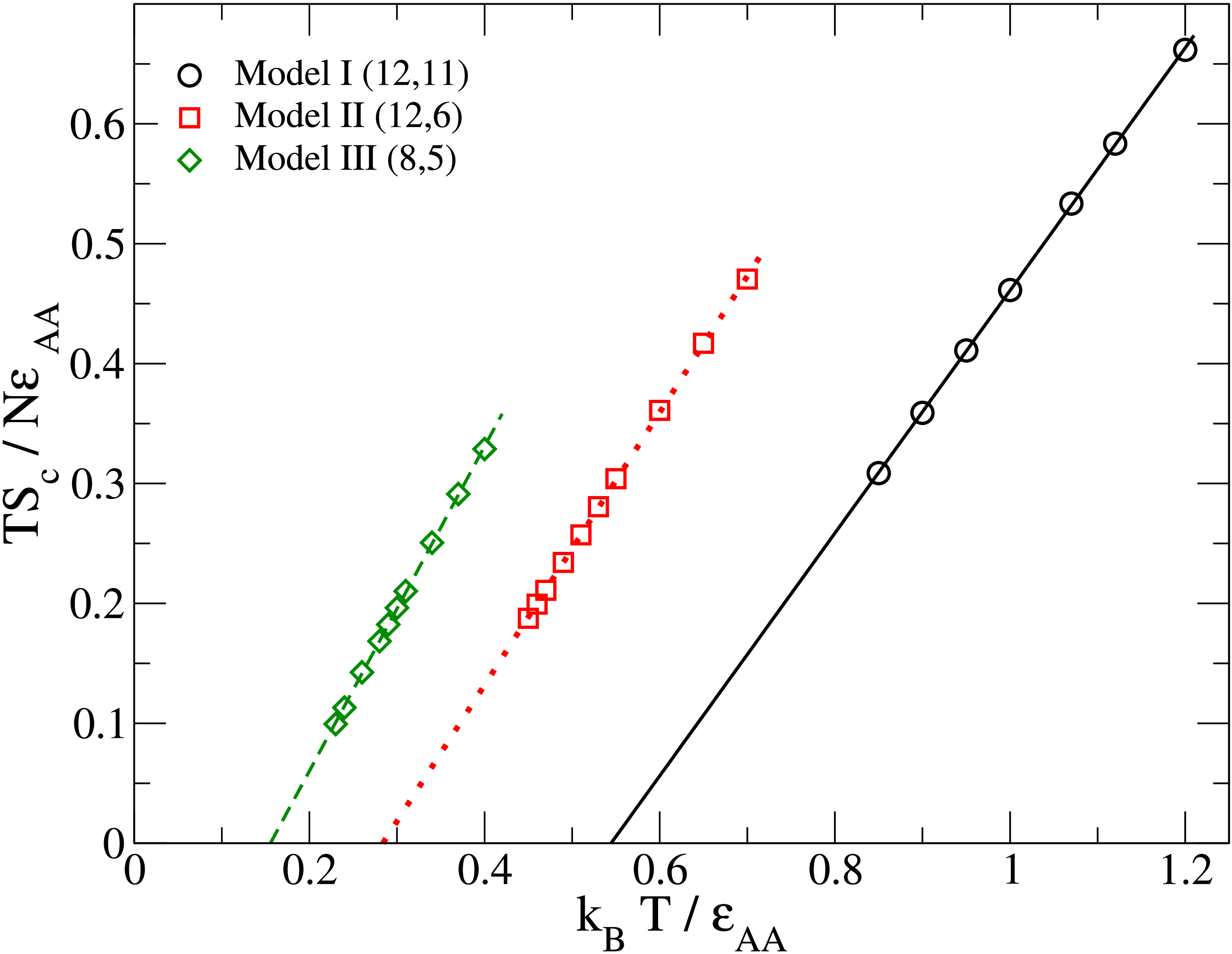}
\caption{Temperature dependence of $TS_{c}$ for the studied models to
  determine the Kauzmann temperature. $T_{K} = 0.54, 0.28, 0.16$
  respectively for models I $(12,11)$, II $(12,6)$, and III $(8,5)$. The value of
  $T_{K}$ from the extrapolated crossing of bulk and basin entropies
  {\it vs.} temperature reported in ~\cite{pap:Sc-Sastry-JPCM} is
  $T_{K}=0.2976$ and in ~\cite{thesis:Karmakar} is $T_{K} \sim
  0.29$. The $T_{K}$ values obtained from this plot is used to
  determine the thermodynamic fragility in
  Fig. \ref{Fig:fragility-allmodel}.}\label{fig:Tsc-vs-T-allmodel}
\end{center}
\end{figure}

\begin{figure}[]
\begin{center}
\includegraphics[width=8cm,height=6cm]{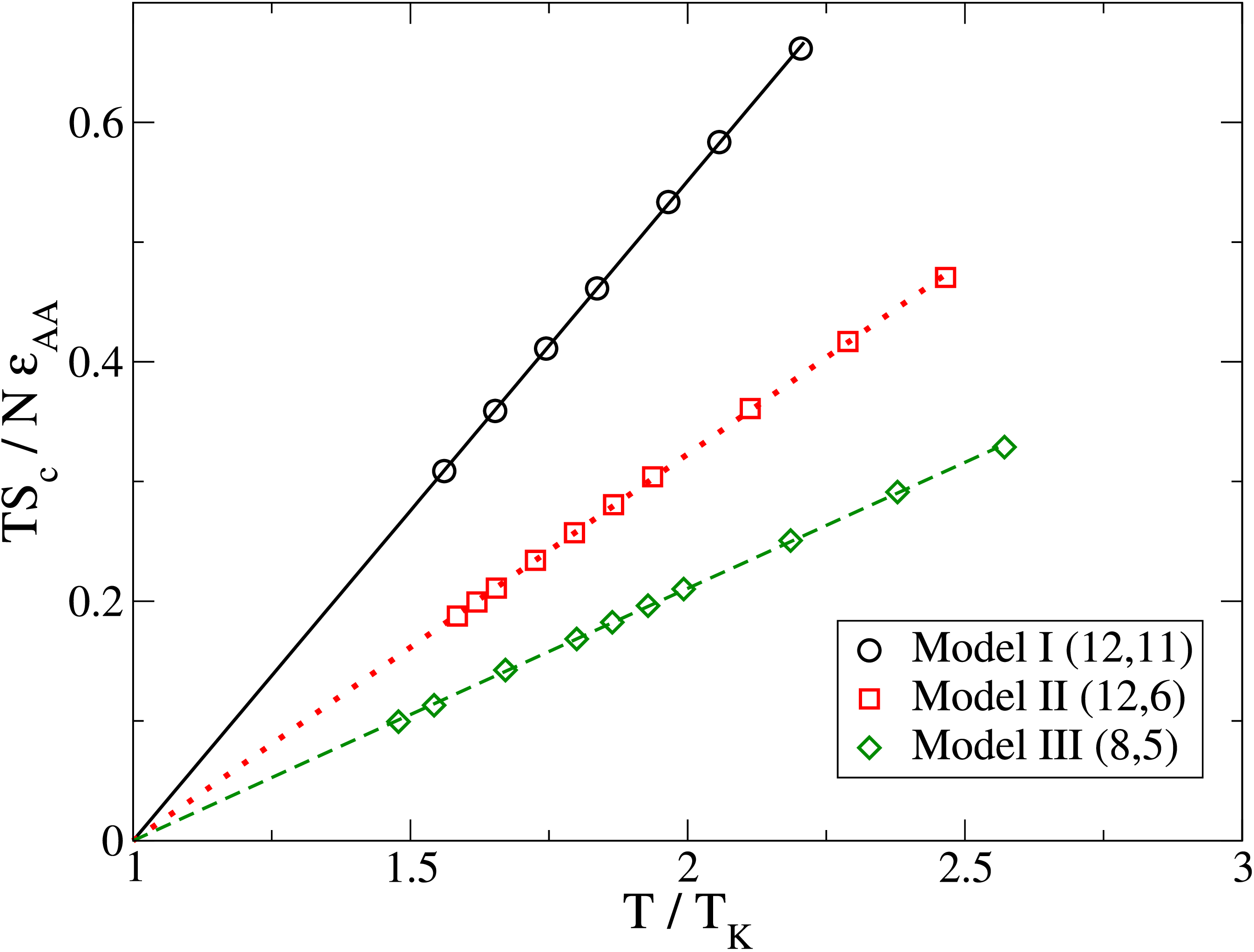}
\caption{Determination of the thermodynamic fragility from the relation
  $TS_{c} = K_{T} (\frac{T}{T_{K}} -1)$ for the studied models. $K_{T}$
  is the slope of the linear fit shown. $T_{K}$ is the
  temperature at which $S_{c}=0$, obtained from the linear fit
  shown in Fig. \ref{fig:Tsc-vs-T-allmodel}. Thermodynamic fragility
  ($K_{T}$) values are $0.551, 0.323, 0.211$ for models I $(12,11)$,
  II $(12,6)$, and III $(8,5)$ respectively}\label{Fig:fragility-allmodel}
\end{center}
\end{figure}

\subsection{Configuration entropy}\label{sec:pel}
Configuration entropy ($S_{c}$) per particle, the measure of the number of distinct local energy minima, is calculated \cite{pap:Sc-Sastry} by subtracting from the total entropy of the system the ‘‘vibrational’’ component:
\begin{equation}
S_{c͑}(T) =  S_{total}(T) - S_{vib} (T)\label{eqn:Sc}
\end{equation}
The total entropy of the liquid is obtained via thermodynamic integration from the ideal gas limit. Vibrational entropy is calculated by making a harmonic approximation to the potential energy about a given local minimum \cite{pap:Sc-Sastry, pap:Sc-Sastry-JPCM, pap:PEL-Sciortino, pap:PEL-Heuer}. The procedure used for generating local energy minima, and calculating the vibrational entropy is as outlined in \cite{pap:Sc-Sastry, pap:Sc-Sastry-JPCM}. 

We have also computed the configuration entropy density $S_{c}(e_{IS}) = k_{B} \ln \Omega(e_{IS})$ where $\Omega(e_{IS})$ is the number density of inherent structures with energy $e_{IS}$ and to a good approximation may be described by a Gaussian.
Equivalently, $S_{c}(e_{IS})$ can be described by a parabola
\begin{equation}
S_{c}(e_{IS}) = \alpha - \frac{(e_{IS} - e_{IS}^{0})^{2}}{\sigma^{2}}
\label{eqn:SceIS}
\end{equation}
The parameter $\alpha$ denotes the peak value of $S_{c}(e_{IS})$ which occurs at energy $e_{IS}^{0}$. $S_{c}(e_{IS})$ is zero 
at $e_{IS} = e_{IS}^{0} \pm \sigma \sqrt{\alpha}$. Thus $\sigma \sqrt{\alpha}$ is a measure of the spread of $S_{c}(e_{IS})$.
We denote the lower root $e_{IS}^{0} - \sigma \sqrt{\alpha}$ by $e_{IS}^{min}$.  

In the harmonic approximation to vibrational entropy, the average value of IS energy sampled by a system at a given temperature $<e_{IS}>(T)$ is predicted to be linear in inverse temperature $\beta=1/T$:
\begin{equation}
\langle e_{IS} \rangle (T) = \langle e_{IS} \rangle (\infty) - \frac{\sigma^{2}}{2T}
\end{equation}
where $\langle e_{IS} \rangle (\infty)$ is the extrapolated limiting value of $\langle e_{IS} \rangle$ at high temperatures. These parameters which characterizes the potential energy landscape are tabulated in Table \ref{tab:trends} for different potentials.

\section{Results and Discussion}

The results from the simulations concerning the thermodynamic and kinetic fragility estimates are presented below. 

\subsection{Thermodynamic fragility}

As described in the Introduction, we may define a thermodynamic
fragility $K_T$ as the slope of $TS_{c}(T)$ {\it vs.}
$T/T_K$. Fig. \ref{fig:Tsc-vs-T-allmodel} shows that indeed,
$TS_{c}(T)$ varies linearly with temperature, which allows us to
define $T_K$. The $T_K$ values for the different potentials are listed
in Table \ref{tab:TempScale}. Various quantities related to the distribution of inherent
structure energies are listed in Table II for later use. Thermodynamic
fragility $K_{T}$ as defined in Eq. ~\ref{eqn:pelfrag} is computed
from the slope of $TS_{c}$ {\it vs.}  $T/T_K$ is found to
\emph{decrease} as the softness of the interaction potential
\emph{increases}, as shown in Fig. \ref{Fig:fragility-allmodel}. Such
behavior is in line with expectations, {\it e. g.} from
\cite{pap:Mattsson-etal,pap:Dudowicz-JPCB-2005}.

\subsection{Kinetic fragility}

\begin{figure*}[t!]
\begin{center}
\subfigure{
\includegraphics[width=7.7cm,height=5.8cm]{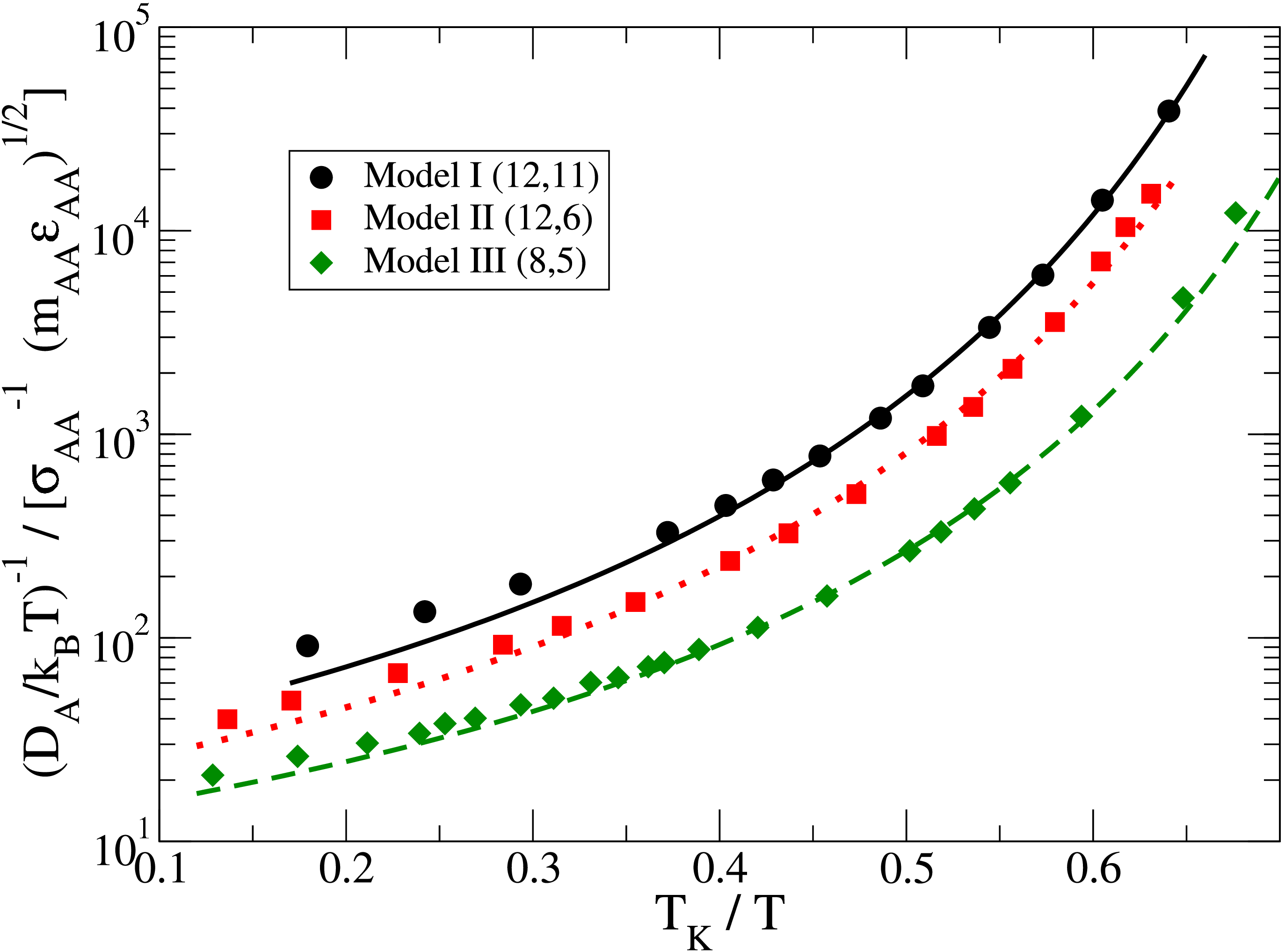}
}
\hspace{2mm}
\subfigure{
\includegraphics[width=7.7cm,height=5.8cm]{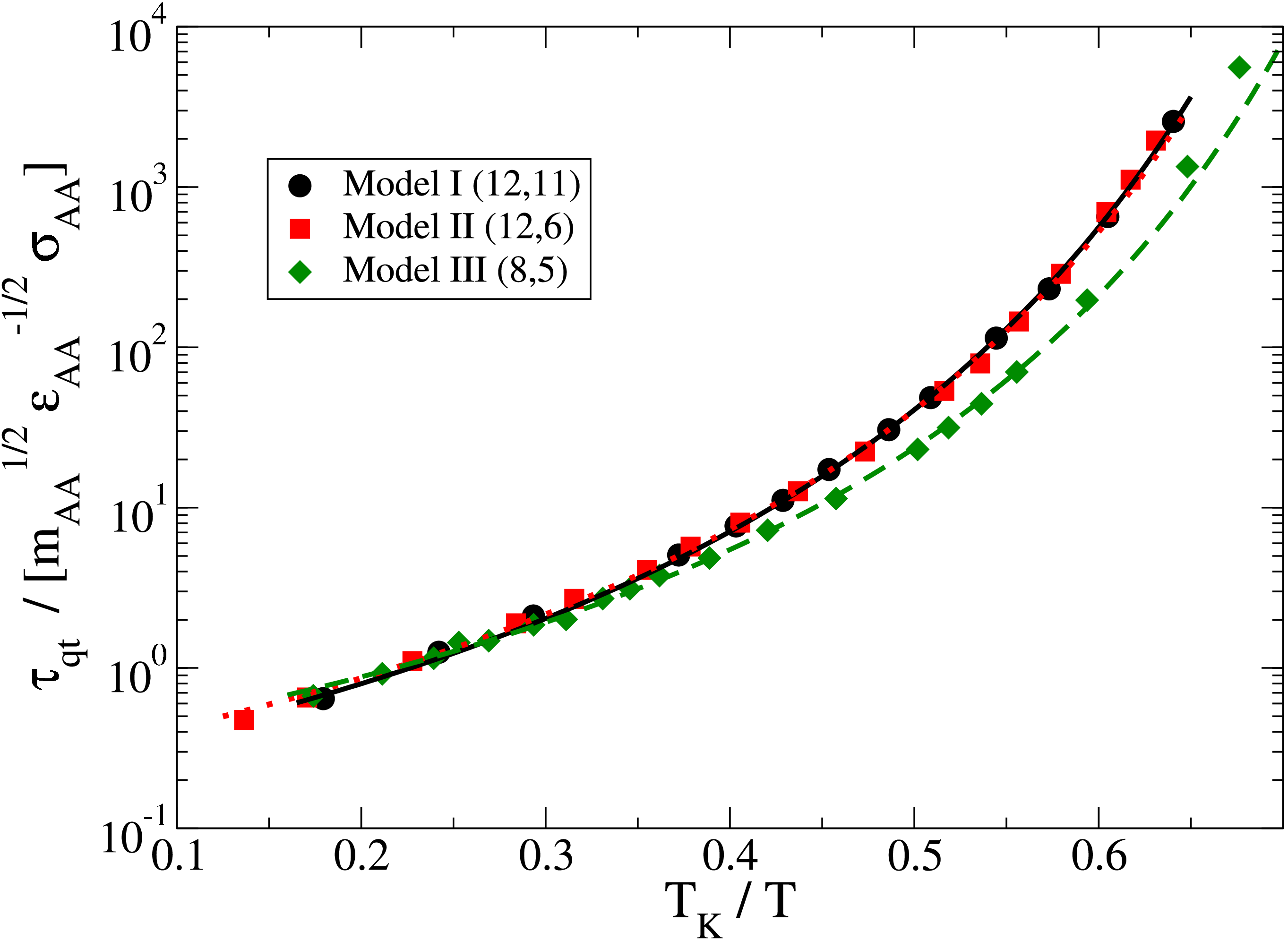}
}
\subfigure{
\includegraphics[width=7.7cm,height=5.8cm]{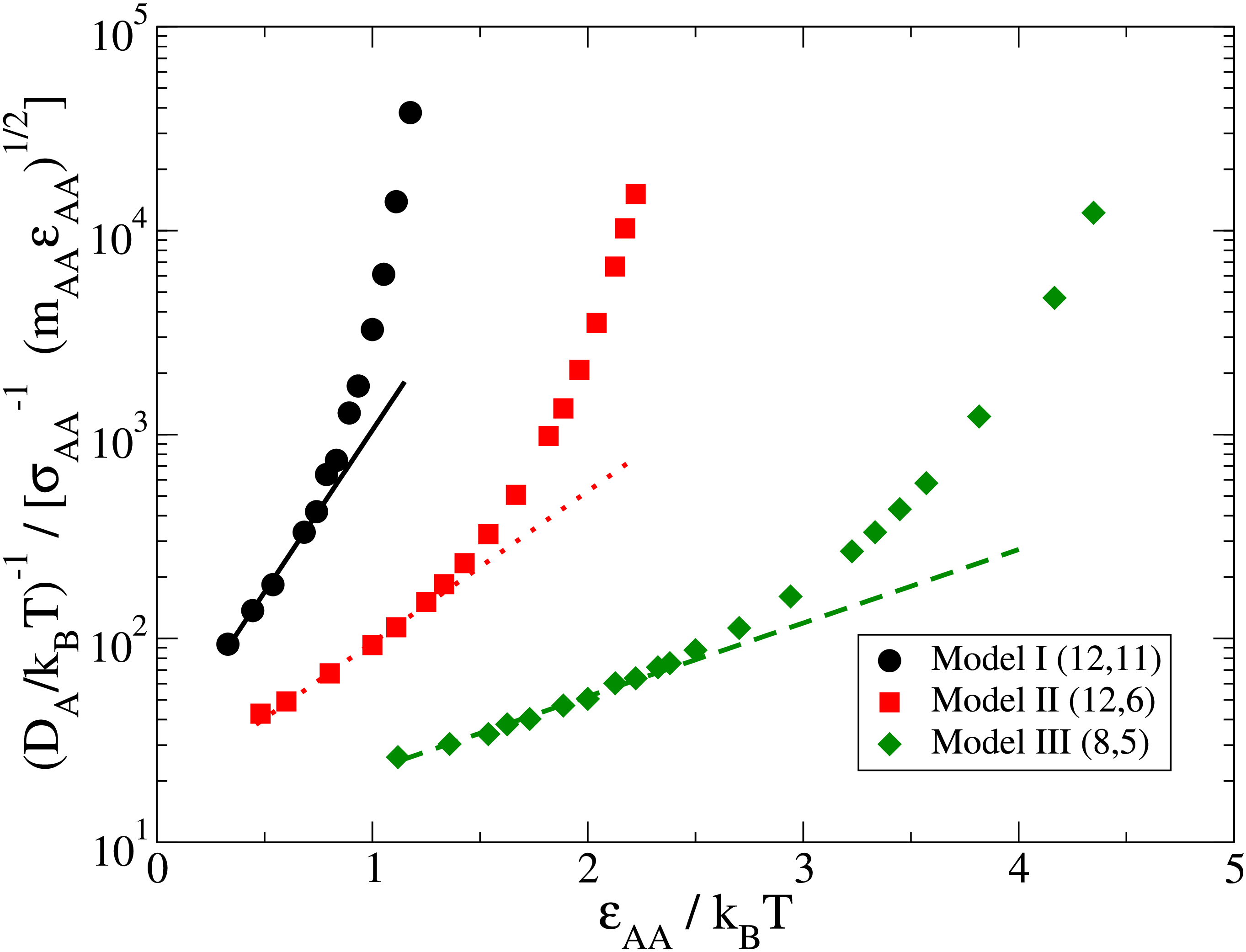}
}
\hspace{2mm}
\subfigure{
\includegraphics[width=7.7cm,height=5.8cm]{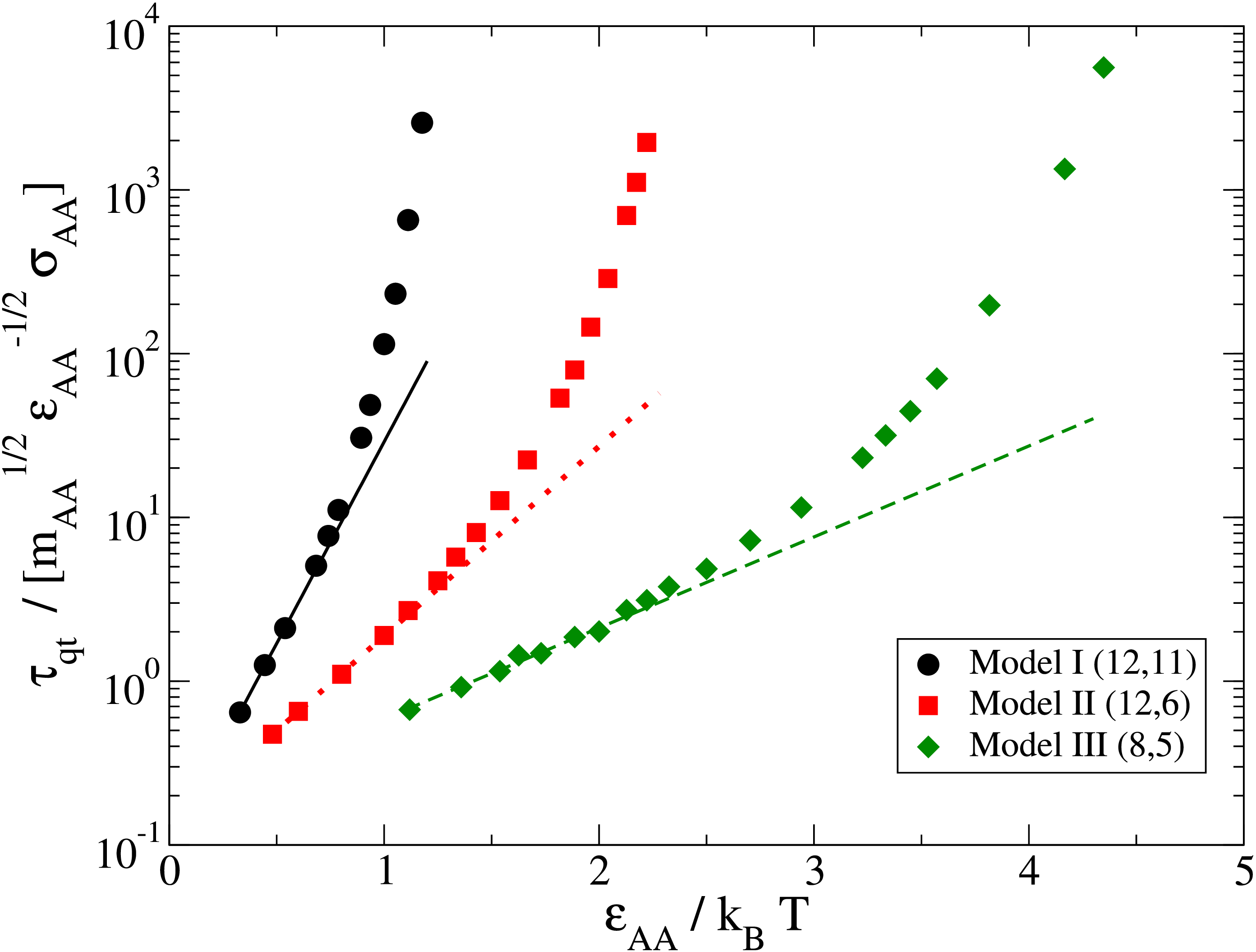}
}
\subfigure{
\includegraphics[width=7.7cm,height=5.8cm]{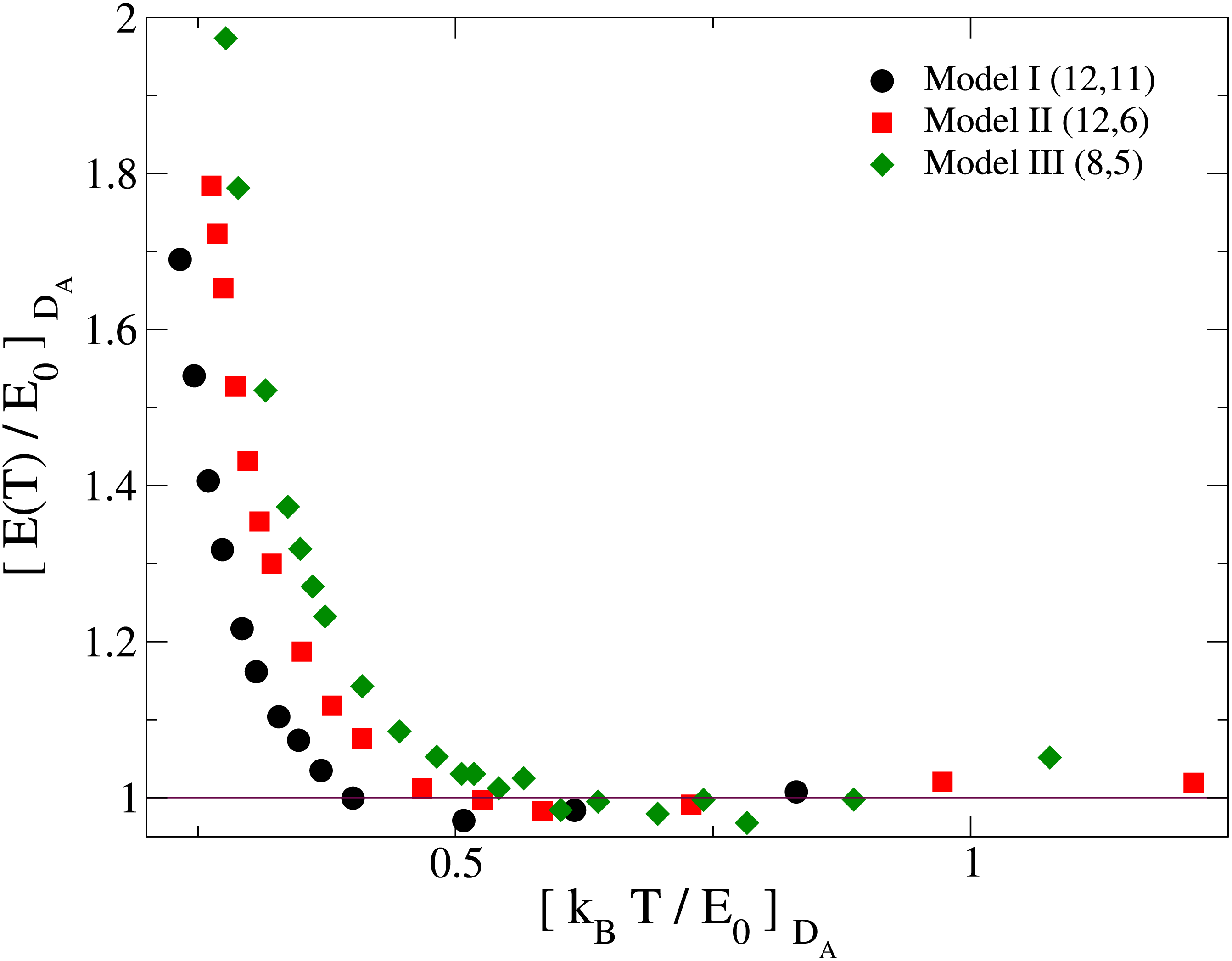}
}
\hspace{2mm}
\subfigure{
\includegraphics[width=7.7cm,height=5.8cm]{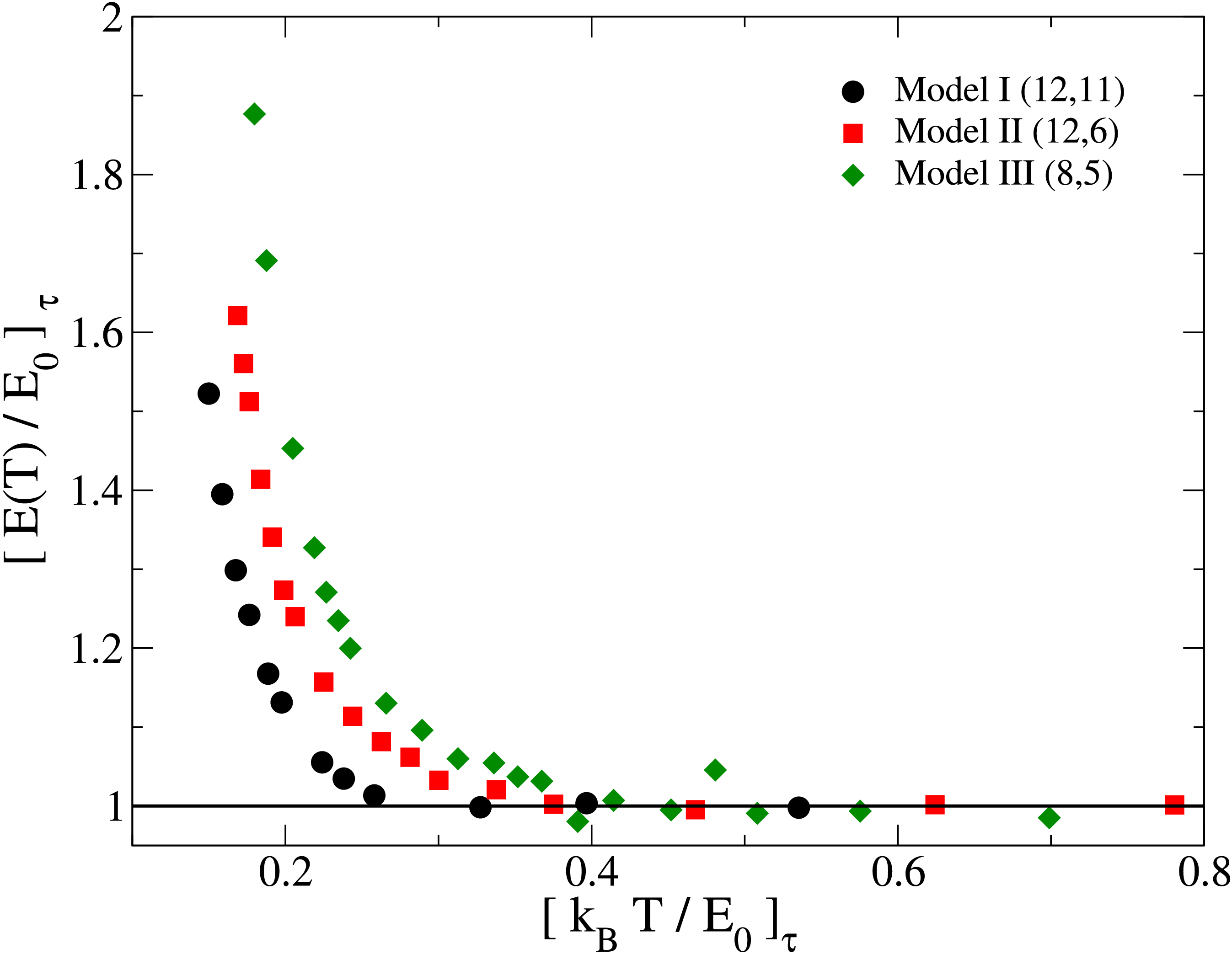}
}
\caption{Top row: Inverse diffusion coefficient and relaxation time
  from overlap function {\it vs.} scaled inverse temperature
  $\frac{T_{K}}{T}$. Lines through the data show VFT fits to the data below the onset temperature. $T_{K}$ estimated from
  fig. \ref{fig:Tsc-vs-T-allmodel} are used as the divergence 
  temperatures in the VFT fits. Middle row: Arrhenius fits to high temperature data
of inverse diffusion coefficient and relaxation time from overlap
  function to determine high temperature activation energies
  $E_{0}$. 
Bottom row: effective activation energy $E(T)$ (see text)
scaled by $E_0$, plotted against $k_B T/E_0$.
}
\label{fig:allmodel-DA-VFTfit}
\end{center}
\end{figure*}

Kinetic fragility ($K_{VFT}$) is estimated by fitting to the VFT form
Eq. ~\ref{eqn:kinfr2} the diffusion coefficients and relaxation times. 

In Fig ~\ref{fig:allmodel-DA-VFTfit} (top panels), we show the
Arrhenius plot of the diffusion coefficients and relaxation times from
$q(t)$, plotted against $T_K/T$. The VFT divergence temperatures $T_{VFT}$, obtained 
from VFT fits to the data for temperatures below the onset
temperature, are found to be
close to $T_K$ and are listed in Table I. The middle panels of Fig
~\ref{fig:allmodel-DA-VFTfit} show Arrhenius fits to high temperature
data (above the onset temperature), from which activation energies
$E_0$ (such that $\tau(T) = \tau_0 \exp(E_0/k_B T)$ are
obtained. These are listed in Table III, and will be discussed later. 
In the bottom panels of Fig ~\ref{fig:allmodel-DA-VFTfit}, we show the 
effective activation energy defined as $E(T) \equiv k_B T \ln( \tau(T)/\tau_{0})$
scaled by $E_0$ (similarly for $D_A/T$), plotted against $k_B T/E_0$.    

We note in the passing that for model $(12,6)$ the proportionality $E_{0} \sim 6 T_{c}$
\cite{pap:Dudowicz-JPCB-2005} is reasonably well satisfied. However, the ratio $E_{0} / T_{c}$ decreases from $\sim 7$ to $\sim 5$ as softness increases.

Next, we calculate the kinetic fragilities $K_{VFT}$, from diffusion
coefficients and relaxation times, using the divergence
temperature $T_{VFT}$ obtained with $T_{VFT}$ as a fit parameter, as well as
using $T_K$ estimates from the configuration entropy as the
divergence temperatures. The corresponding kinetic fragilities,
labeled $K_{VFT}^{I}$ and $K_{VFT}^{II}$, are listed in Table IV,
along with the thermodynamic fragilities $K_T$. We find that the
kinetic fragilities \emph{increase} as the softness of the interaction
potential \emph{increases}, thus showing a trend that is opposite to
that of the thermodynamic fragility.

\subsection{Adam Gibbs relation and fragility}

\begin{figure*}[!]
\begin{center}
\includegraphics[width=8cm,height=6cm]{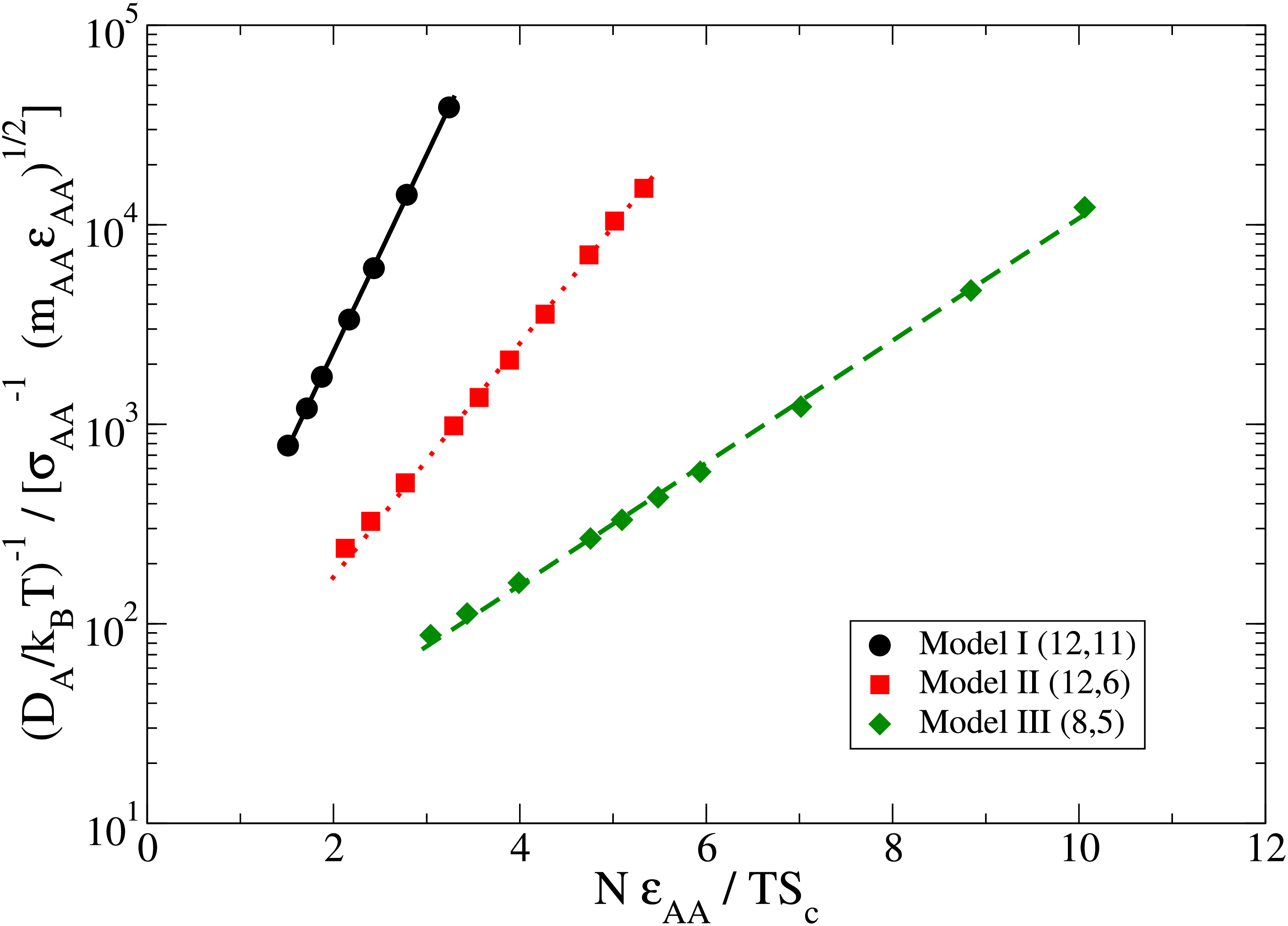}
\hspace{2mm}
\includegraphics[width=8cm,height=6cm]{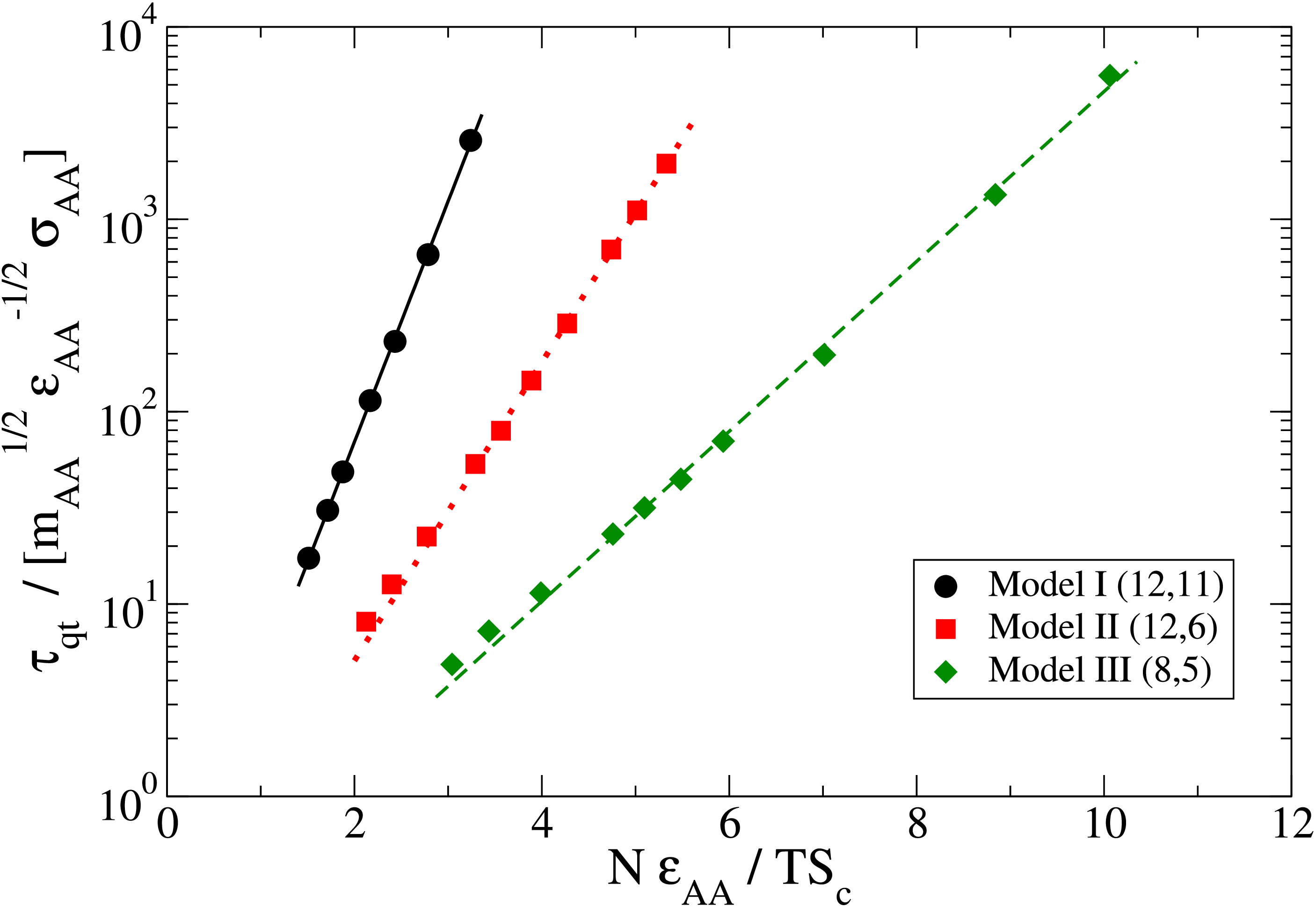}
\caption{Adam Gibbs plots for the inverse diffusion coefficient of $A$
 particles and relaxation time from overlap function, for the three models studied. The activation
 energy parameter $A$ in Eq. \ref{eqn:AG}, obtained from the slopes of the data shown, is tabulated in table
 \ref{tab:energies}. } \label{Fig:ag-allmodel}
\end{center}
\end{figure*}

In order to understand this discrepancy, we consider again the
Adam-Gibbs relation, which relates the kinetic and thermodynamic
fragilities. Comparing Eq. \ref{eqn:kinfr2}, Eq. \ref{eqn:AG} and
Eq. \ref{eqn:pelfrag}, we note that the relationship between the
kinetic and thermodynamic fragilities that we may deduce assuming the validity 
of the VFT and the AG relations is 
\begin{equation} 
K_{VFT} = K_{T}/A 
\end{equation} 
and we expect at least the same trend in the two fragilities under the
assumption that the term $A$ does not substantially alter the
proportionality between kinetic and thermodynamic fragilities. To
assess the degree to which this is true in our models, we show in
Fig. \ref{Fig:ag-allmodel} the Adam-Gibbs plots of the diffusion coefficient and
relaxation times. These plots show that the coefficient $A$, obtained
from the slopes (and listed in Table III), indeed varies from one
model to the other, \emph{decreasing} as the softness increases. Thus,
the ratio $K_{T}/A$ shows the opposite trend, \emph{increasing} as the
softness increases. 

\begin{figure}[!]
\begin{center}
\includegraphics[scale=0.25]{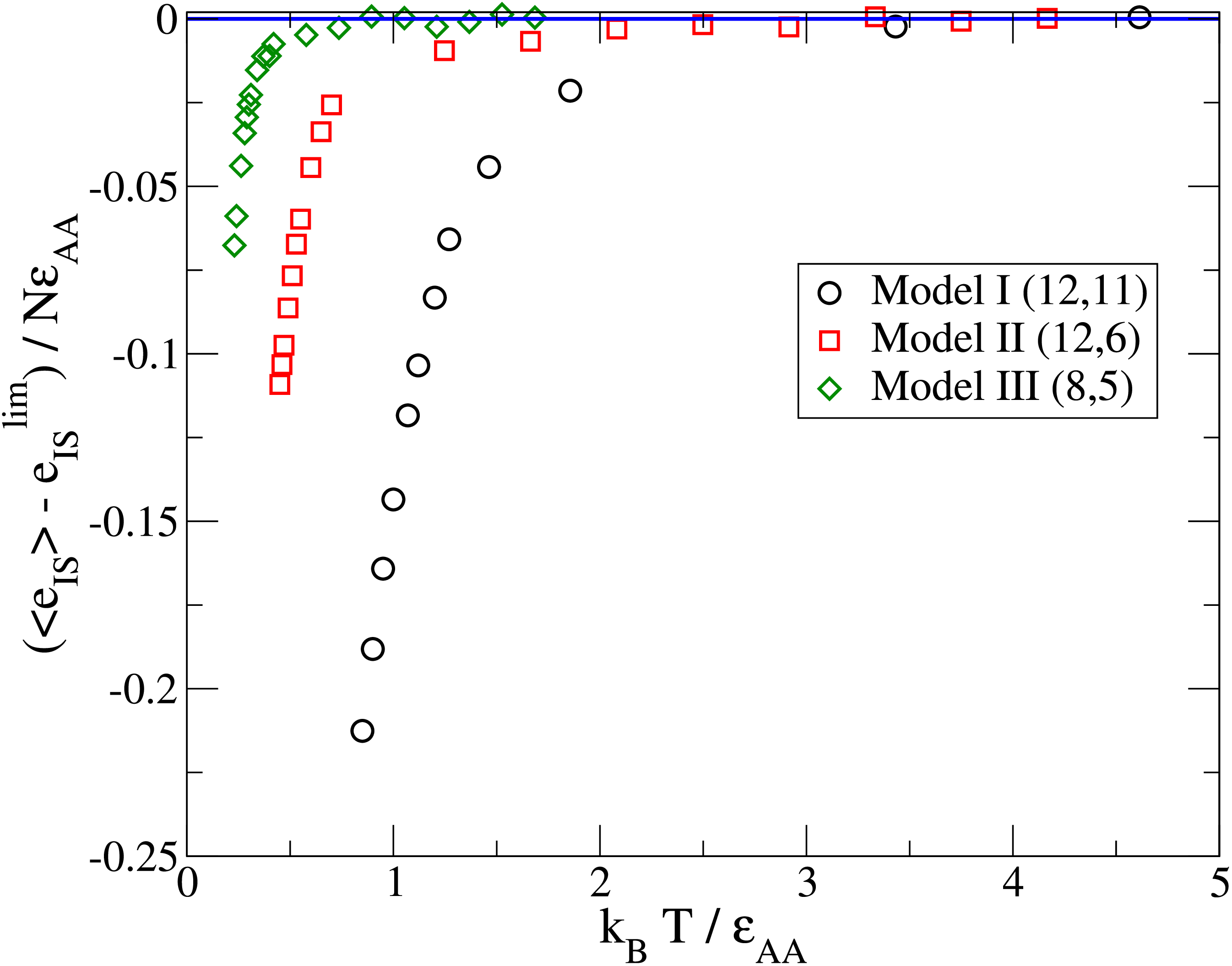}
\caption{Temperature dependence of the inherent structure energy $e_{IS}$ for the studied models shifted by the corresponding high temperature limiting values $e_{IS}^{lim}$ for clarity. The values of $e_{IS}^{lim}$ are $-6.003, -6.886, -7.191$ for models I $(12,11)$, II $(12,6)$ and III $(8,5)$ respectively.}\label{Fig:eis-limittingvalue-allmodel}
\end{center}
\end{figure}

We next attempt to understand the dependence of the Adam-Gibbs
coefficient $A$ on the softness of the interaction. First we consider
the high temperature Arrhenius behavior of relaxation times, in terms
of the Adam-Gibbs relation. Such Arrhenius behavior can be expected if
the configuration entropy effectively becomes a constant, in which
case, the high temperature activation energy will be given by 
\begin{equation}
E_0 = A k_B/Sc_{\infty}
\label{eqn:E0-AbySc}
\end{equation}
However, the asymptotic high temperature configuration entropy is
difficult to assess directly, as the various available approaches to
computing the basin entropy do not work well in this regime(see {\it
  e. g.} \cite{pap:Sastry-pcc}). We thus use the following procedure:
First, we determine directly from simulations the high temperature
limit of the inherent structure energies, $e_{IS}^{lim}$ (see Fig. \ref{Fig:eis-limittingvalue-allmodel}).
Then, we use
the extrapolation of the dependence of the configuration entropy $S_c$
on the inherent structure energy $e_{IS}$ obtained below the onset
temperature to obtain the high temperature limit of the configuration
entropy, $S_c(e_{IS}^{lim})$, which do not vary appreciably with
softness of interaction, and are listed in Table II. Table II also
lists $S_c(\infty)$, the infinite temperature value of $S_c$ obtained
by extrapolating Eq. \ref{eqn:pelfrag} to infinite temperature, a
procedure that is not justified at temperatures above the onset
temperature.  Using these $S_c(e_{IS}^{lim})$
values, and the activation energies $E_{0}$ shown in Table III, we
obtain estimates for the AG coefficient
\begin{equation}
A_{est} = E_0 S_c(e_{IS}^{lim})/k_B
\label{eqn:Aest-from-E0}
\end{equation}
which are shown in Table III. We note in Table III that $E_0$ values
decrease strongly as the softness of the interactions increases, and
with a corresponding moderate increase of $S_c(e_{IS}^{lim})$, our
estimates of $A_{est}$ agree rather well with the values obtained
directly from the Adam-Gibbs plots. We now designate the thermodynamic
fragility estimates obtained by considering the full form of the
Adam-Gibbs relation as $K_{AG} = K_T/A$, and list them along side the
thermodynamic and kinetic fragility estimates in Table IV. As expected
from the above discussion, the ``Adam-Gibbs'' fragility estimates
($K_{AG}^{I}$ in Table IV) agree rather well with the kinetic fragilities. 

\begin{figure*}[!]
\begin{center}
\includegraphics[width=7cm,height=6cm]{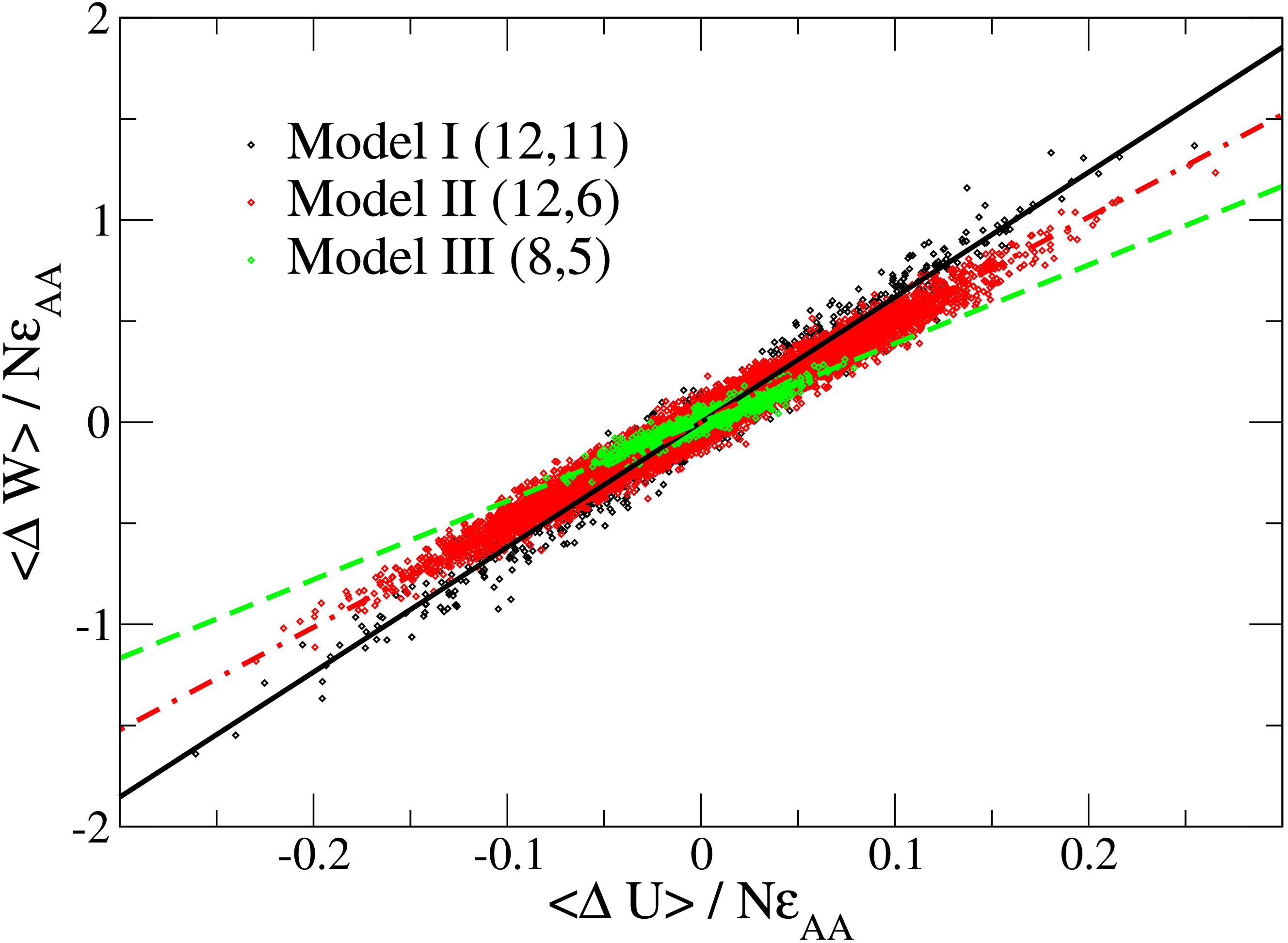}
\hspace{1cm}
\includegraphics[width=7.5cm,height=6cm]{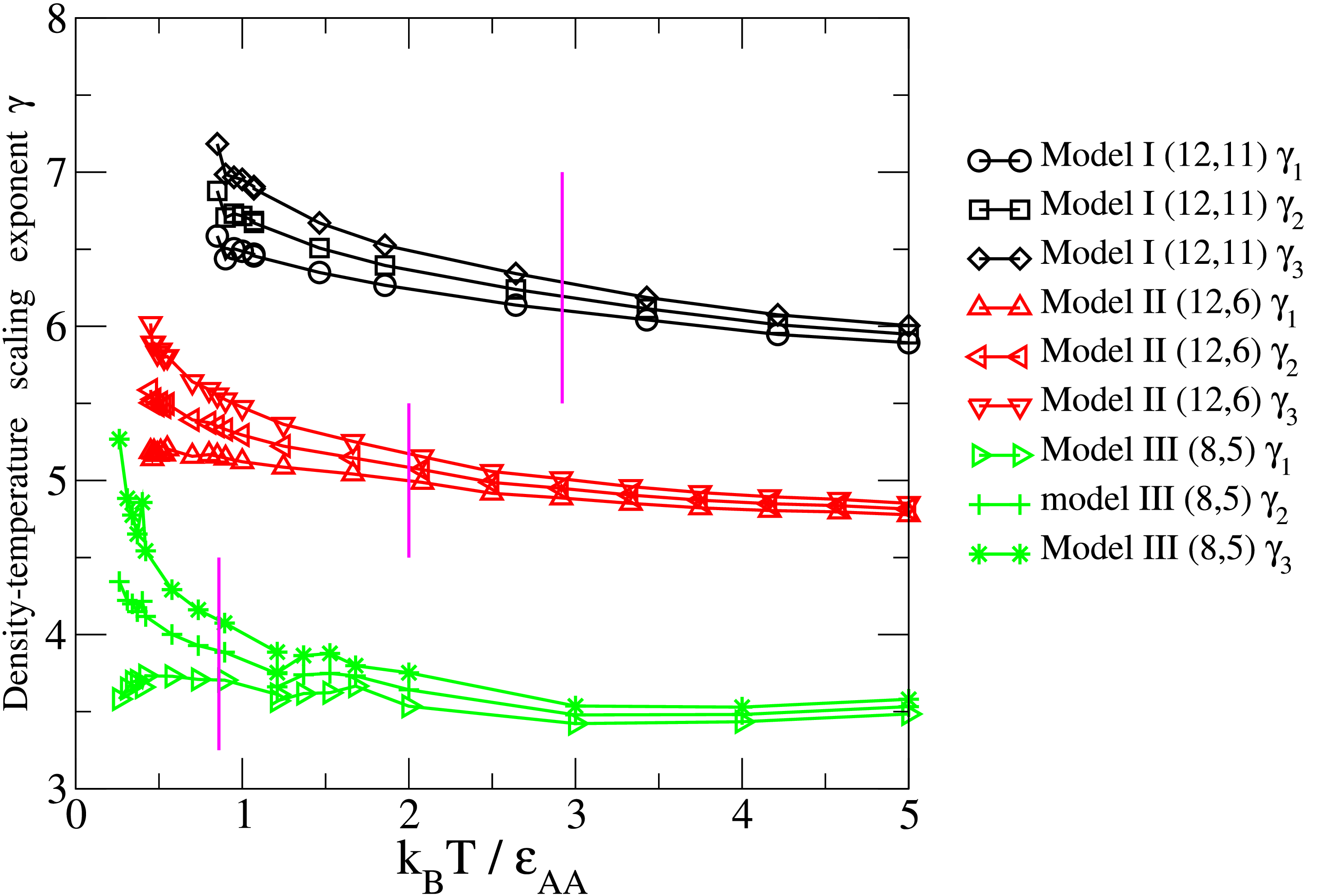}
\caption{Determination of the density-temperature scaling exponent
  $\gamma$ from the correlation between instantaneous potential energy
  ($U$) and virial ($W$). $\gamma_{1} = \frac{\langle \Delta W \Delta
    U \rangle} {\langle (\Delta U)^{2} \rangle}$, $\gamma_{2} =
  \frac{\sqrt{\langle (\Delta W)^{2}}}{\sqrt{\langle (\Delta U)^{2}
      \rangle}}$, $\gamma_{3} = \frac{\langle (\Delta W)^{2}
    \rangle}{\langle \Delta W \Delta U \rangle}$ where $\Delta U = U -
  \langle U \rangle$ and $\Delta W = W - \langle W \rangle$ represent
  fluctuations about mean of potential energy and virial
  respectively. The left panel shows the correlation between energy and
  virial at temperatures $\approx 2 T_{onset}$, with straight line 
fits $\langle \Delta W \rangle =
  \gamma_{2}(2T_{onset}) \langle \Delta U \rangle$. The right panel shows the temperature dependent values of $\gamma$ for the studied models.} \label{Fig:allmodel-fluctuation}
\end{center}
\end{figure*}

Although the above picture provides a consistent description of the
fragilities from kinetic and thermodynamic data, a question remains
regarding the variation of the high temperature activation energy
$E_0$ with the softness of the interaction potential. To seek some
insight into this question, we consider work in recent years
concerning the scaling of the temperature dependence of dynamic and
thermodynamic quantities at different densities
\cite{schroder,tarjus,roland}. It has been shown by many groups that a        
scaled variable $\rho^{\gamma}/T$, where $\rho$ is the density,
captures the density variation of properties in many liquids. The
exponent $\gamma$ can easily be shown to be $n/3$ for inverse power law
potentials, where $n$ is the power of the inverse power law, but even
for other liquids, an effective $\gamma$ has been shown to be
derivable by considering the correlated fluctuations of potential
energy and the virial \cite{schroder}. The exponent $\gamma$ is              
obtainable as the ratio of fluctuations.
Although such a ratio is state point dependent, a ``best
fit'' value, typically obtained from high temperature state points,
has been shown to effectively describe the scaling of properties at
different densities. Since we do not perform a full analysis of the
density dependence here, we do not estimate the best value of $\gamma$
but instead use the value at twice the onset temperature as an
indicative value. Fig. \ref{Fig:allmodel-fluctuation} shows the fluctuation data from which the
$\gamma$ value is obtained, and the temperature variation
of the exponents. The values of $\gamma$ we use are shown in Table
II.

Based on the above considerations, we should expect the high
temperature activation energies to be proportional to $\rho^{\gamma}$.
Accordingly, we obtain estimates of the activation energy in the form
$E_0 = E_{00} \rho^{\gamma}$. These values, shown in Table III, have a
weaker temperature dependence than the directly evaluated $E_0$, and
correspondingly, the fragility estimates obtained (shown in Table IV),
while showing a smaller decrease with softness, nevertheless decrease
with increasing softness of interaction. A further analysis is needed,
therefore, to elucidate the relevance of these considerations to
evaluating the variation of the high temperature activation energy.

\section{Conclusions}

We have studied the effect of the softness of the interaction
potential on fragility in three model glass formers. We find that the
kinetic fragility obtained from diffusion coefficients and relaxation
times \emph{increases} with increasing softness of the interaction
potential, contrary to expectations based on earlier studies
\cite{pap:Mattsson-etal,pap:Dudowicz-JPCB-2005}. On the other hand, a thermodynamic fragility
obtained from the temperature variation of the configuration entropy
\emph{decreases} with increasing softness of the interaction
potential.  By taking into consideration the model dependence of the
high temperature activation energy, in addition to the temperature
dependence of the configuration entropy, we define an ``Adam-Gibbs''
fragility whose model dependence accurately captures the variation of
the kinetic fragilities that we find. An attempt to rationalize the
model dependence of the high temperature in terms of the scaling of
properties with respect to density is encouraging but fails to fully
explain the observed decrease of the fragility with increasing
softness of the interaction potential.

\begin{acknowledgments}
We would like to thank Thomas B. Schr\o der and Jack Douglas for critical reading of the manuscript. We thank CCMS, JNCASR for computational facilities. S. Sengupta thanks CSIR for financial support.
\end{acknowledgments}

\begin{table*}
\caption{\label{tab:trends} Potential energy landscape parameters and density temperature scaling exponents for the studied models. Fit forms used: $\langle e_{IS} \rangle (T) = \langle e_{IS} \rangle (\infty) - \frac{\sigma^{2}}{2T}$; $S_{C}(e_{IS}) = \alpha - \frac{(e_{IS} - e_{IS}^{0})^{2}}{\sigma^{2}}$. }
\begin{ruledtabular}
\begin{tabular}{llll}
Quantity & (12,11) & (12,6) & (8,5)\\
\hline
Density minimum for IS pressure & 1.04 & 1.09 & 1.18\\
\\
Height of $S_{C}(e_{IS})$ distribution $\alpha$            & 0.863 &  0.886 & 0.905\\
Spread of $S_{C}(e_{IS})$ distribution $\alpha^{1/2}\sigma$ & 0.816 &  0.455 & 0.255\\
IS Energy where $S_{C}(e_{IS})=0$, $e_{IS}^{min}= e_{IS}^{0} - \sigma\sqrt{\alpha}$  & -6.457 & -7.132 &  -7.346\\
$\langle e_{IS} \rangle (\infty)$ & -5.761 & -6.734 & -7.098\\
Limiting value of IS energy $e_{IS}^{lim}$ & -6.003 & -6.886 & -7.191\\
$S_{c}(e_{IS}^{lim})$ & 0.69 & 0.7 & 0.78\\
$S_{c}(\infty) = \frac{K_{T}}{T_{K}}$ &  1.01 & 1.14 &  1.35\\
\\
Density temperature scaling exponent at $2 \times T_{onset}$ & & &\\
$\gamma_{1}$  & 6.09 & 4.99 & 3.71\\
$\gamma_{2}$  & 6.18 & 5.07 & 3.89\\ 
$\gamma_{3}$  & 6.27 & 5.15 & 4.09\\ 
$\rho^{\gamma_{1}}=1.2^{\gamma_{1}}$ & 3.04 & 2.48 & 1.97\\  
$\rho^{\gamma_{2}}=1.2^{\gamma_{2}}$ & 3.09 & 2.52 & 2.03\\ 
$\rho^{\gamma_{3}}=1.2^{\gamma_{3}}$ & 3.14 & 2.56 &  2.11\\
\\
\hline
\end{tabular}
\end{ruledtabular}
\end{table*}

\begin{table*}
\caption{\label{tab:energies} Comparison of activation energy
  parameters. $A$ is the activation parameter in the Adam Gibbs (AG)
  relation. $E_{0}$ is the high temperature activation energy in
  Arrhenius fit.  $S_{c}(e_{IS}^{lim})$ is the values of configuration
  entropy density at the limiting value of inherent structure energies
  at high temperatures. $A_{est}^{I} = E_{0}S_{c}(e_{IS}^{lim})/k_{B}$
  is the expected value of parameter $A$ obtained from
  $E_{0}$. $E_{0}^{est}=E_{00} \rho^{\gamma}$ is the estimate of
  $E_{0}$ from density - temperature scaling of relaxation
  time where $\gamma_{2}$  are the values of the scaling exponent at twice the onset temperature. $A_{est}^{II} =
  E_{0}^{est}S_{c}(e_{IS}^{lim})$ is the expected value of energy
  barrier $A$ obtained from $E_{0}^{est}$. }
\begin{ruledtabular}
\begin{tabular}{l|c|cc|ccc|c|cc|ccc}
 & \multicolumn{6}{c|}{From $q(t)$} &       \multicolumn{6}{c}{From $(\frac{D_{A}}{T})^{-1}$}\\
\hline
\multirow{2}{*}{Model} & & & & & & & & & & & & \\
    &$A$&$E_{0}$&$A_{est}^{I}$&$E_{00}$&$E_{0}^{est}$&$A_{est}^{II}$&$A$&$E_{0}$&$A_{est}^{I}$&$E_{00}$&$E_{0}^{est}$&$A_{est}^{II}$\\
\hline 
(12,11) & 2.88 & 5.67 & 3.91 &      & 4.13 & 2.85 & 2.27 & 3.65 & 2.52 &      & 2.63 & 1.81\\ 
(12,6)  & 1.79 & 2.67 & 1.87 & 1.34 & 3.38 & 2.40 & 1.35 & 1.71 & 1.20 & 0.85 & 2.14 & 1.52\\ 
(8,5)   & 1.02 & 1.28 & 1.00 &      & 2.72 & 2.12 & 0.71 & 0.83 & 0.65 &      & 1.73 & 1.35\\ 
\hline
\end{tabular}
\end{ruledtabular}
\end{table*}

\begin{table*}
\caption{\label{tab:fragilites} Comparison of fragility
  parameters. $K_{T}$ is thermodynamic fragility obtained from
  temperature dependence of $TS_{c}(T)$. $K_{VFT}^{I}$ is kinetic
  fragility from VFT fit and $K_{VFT}^{II}$ is kinetic fragility
  obtained from VFT fit assuming $T_{VFT}=T_{K}$. $K_{AG}^{I}=\frac{K_{T}}{A_{est}^{I}}$ is the
  fragility expected from high temperature activation energy $E_{0}$
  obtained from an Arrhenius
  fit. $K_{AG}^{II}=\frac{K_{T}}{A_{est}^{II}}$ where the high temperature
  activation energy is estimated from density-temperature scaling. }
\begin{ruledtabular}
\begin{tabular}{l|c|cc|cc|cc|cc}
    & & \multicolumn{4}{c|}{From $q(t)$} &       \multicolumn{4}{c}{From $(\frac{D_{A}}{T})^{-1}$}\\
\hline
\multirow{2}{*}{Model} & & & & & & & & & \\
    &  $K_{T}$ &$K_{VFT}^{I}$&$K_{VFT}^{II}$&$K_{AG}^{I}$& $K_{AG}^{II}$ &$K_{VFT}^{I}$ &$K_{VFT}^{II}$ &$K_{AG}^{I}$ &$K_{AG}^{II}$\\
\hline
(12,11) &  0.551 & 0.20 & 0.19 & 0.14 & 0.19 & 0.34 & 0.24 & 0.22 & 0.30\\
(12,6)  &  0.323 & 0.21 & 0.20 & 0.17 & 0.13 & 0.38 & 0.26 & 0.27 & 0.21\\
(8,5)   &  0.211 & 0.26 & 0.23 & 0.21 & 0.10 & 0.40 & 0.32 & 0.32 & 0.16\\ 
\hline 
\end{tabular}
\end{ruledtabular}
\end{table*}


\begin{thebibliography}{}

\bibitem{fragility_angell} C. A. Angell, {\it J. Non-Cryst. Solids} {\bf 131-133}, 13 (1991); R. B\"{o}hmer, K. L. Ngai, C. A. Angell and D. J. Plazek, {\it J. Chem. Phys.} {\bf 99}, 4201 (1993); C. A. Angell, {\it Science}, {\bf 267}, 1924 (1995); 


\bibitem{speedy} R. J. Speedy, {\it J. Phys. Chem. B} {\bf 103}, 4060 (1999).


\bibitem{pap:AG-Sastry}S. Sastry, {\it Nature}   {\bf 409}, 164 (2001).


\bibitem{wales} D. J. Wales and J. P. K. Doye {\it Phys. Rev. B} {\bf 63}, 214204 (2001); vol. 64, 024205 (2001).



\bibitem{tarjus}C. Alba-Simionesco, D. Kivelson, and G. Tarjus, {\it J. Chem. Phys.} {\bf 116}, 5033 (2002); G. Tarjus, D. Kivelson, S. Mossa and C. Alba-Simionesco, {\it J. Chem. Phys.} {\bf 120}, 6135 (2004); C. Alba-Simionesco, A. Cailliaux, A. Alegr\'{i}a  and G. Tarjus, {\it Europhys. Lett.}, {\bf 68}, 58, (2004). 




\bibitem{ruocco}G. Ruocco,  F. Sciortino, F. Zamponi, C. De Michele and T. Scopigno, {\it J. Chem. Phys.}, {\bf 120}, 10666 (2004).




\bibitem{sokolov} V. N. Novikov and A. P. Sokolov, {\it Nature} {\bf 431}, 961 (2004). 



\bibitem{pap:BordatPRL}P. Bordat, F. Affouard, M. Descamps, {\it Phys. Rev. Lett.} {\bf 93}, 105502 (2004).

\bibitem{pap:BordatJNCS}P. Bordat, F. Affouard, M. Descamps, {\it J. Non Cryst. Solids} {\bf 353}, 3924 (2007).




\bibitem{pap:Dudowicz-JPCB-2005}J. Dudowicz, K. F. Freed and J. F. Douglas, {\it J. Phys. Chem. B} {\bf 109}, 21350 (2005).

\bibitem{pap:Dudowicz-JCP-2005}J. Dudowicz, K. F. Freed and J. F. Douglas, {\it J. Chem. Phys.} {\bf 123}, 111102 (2005).


\bibitem{douglas} J. F. Douglas, J. Dudowicz and K. F. Freed, {\it J. Chem. Phys.} {\bf 125}, 144907 (2006); R. A. Riggleman, J. F. Douglas and J. J. de Pablo, {\it J. Chem. Phys.} {\bf 126}, 234903 (2007).





\bibitem{francis} F. W. Starr and J. F. Douglas, {\it Phys. Rev. Lett.} {\bf 106} 115702 (2011).




\bibitem{sneha} S. E. Abraham, S. M. Bhattacharrya, and B. Bagchi, {\it 
Phys. Rev. Lett.} {\bf 100}, 167801 (2008). 


\bibitem{tanaka} H. Shintani and H. Tanaka, {\it Nat. Mater.} {\bf 7}, 870 (2008). 




\bibitem{pap:Mattsson-etal} Johan Mattsson, Hans M. Wyss, Alberto Fernandez-Nieves, Kunimasa Miyazaki, Zhibing Hu, David R. Reichman and David A. Weitz, {\it Nature (London)}, {\bf 462}, 83 (2009).


\bibitem{pap:Angell-news-views}C. A. Angell and K. Ueno, {\it Nature} {\bf 462}, 45, (2009).


\bibitem{AdamGibbs} G. Adam and J. H. Gibbs, {\it J. Chem. Phys.} {\bf 43}, 139 (1965).



\bibitem{Rossler} E. Rossler, K.-U. Hess, V. N. Novikov,
{\it J. Non-Cryst. Solids} {\bf 223}, 207 (1998).

\bibitem{Chandler} Y. S. Elmatad, D. Chandler, and J. P. Garrahan, {\it J  Phys Chem B} {\bf 114}, 17113 (2010); and {\it J. Phys. Chem. B.} {\bf 113}, 5563 (2009).

\bibitem{Dyre} T. Hecksher, A. I. Nielsen, N. B. Olsen And J. C. Dyre, {\it Nature Physics} {\bf 4} 737 (2008).


\bibitem{pap:PEL-Sciortino}F. Sciortino, {\it J. Stat. Mech.} P05015 (2005).

\bibitem{pap:PEL-Heuer}A. Heuer, {\it J. Phys.: Condens. Matter} {\bf 20}, 373101 (2008).


\bibitem{inh} F. H. Stillinger and T. A. Weber, {\it Science} {\bf 225},983 (1984); 
F. H. Stillinger, {\it Science} {\bf 267}, 1935 (1995). 



\bibitem{schroder}Nicoletta Gnan, Thomas B. Schrøder, Ulf R. Pedersen, Nicholas P. Bailey, and Jeppe C. Dyre, {\it J. Chem. Phys.} {\bf 131}, 234504 (2009);   Thomas B. Schrøder, Nicoletta Gnan, Ulf R. Pedersen, Nicholas P. Bailey, and Jeppe C. Dyre,  {\it J. Chem. Phys.} {\bf 134}, 164505 (2011) and other papers in the series.


\bibitem{pap:Sc-Sastry}S. Sastry,  {\it Phys. Rev. Lett.}  {\bf 85}, 590 (2000).

\bibitem{pap:Sc-Sastry-JPCM}S. Sastry, {\it J. Phys.: Condens. Matter} {\bf 12}, 6515 (2000).


\bibitem{kob} W. Kob and H. C. Andersen, Phys. Rev. E 51, 4626 (1995).


\bibitem{pap:BC} D. Brown and J. H. R. Clarke, {\it Mol. Phys.} {\bf 51}, 1243 (1984).


\bibitem{pap:4pt-CD}C. Dasgupta,  A. V. Indrani, S. Ramaswamy and M. K. Phani, {\it Europhys. Lett.} {\bf 15}, 307 (1991).

\bibitem{pap:Ovlap-Glotzer-etal}S. C. Glotzer, V. N. Novikov and T. B. Schr\o der, {\it J. Chem. Phys.} {\bf 112}, 509 (2000).

\bibitem{pap:Lacevic}N. La\u{c}evi\'{c}, F. W. Starr, T. B. Schr\o der, and S. C. Glotzer, {\it J. Chem. Phys.} {\bf 119}, 7372 (2003).

\bibitem{pap:Ovlap-Donati-etal}C. Donati, S. Franz, S. C. Glotzer and G. Parisi,  {\it J. Non-Cryst Solids} {\bf 307}, 215–224 (2002).

\bibitem{pap:Karmakar-PNAS}S. Karmakar, C. Dasgupta, S. Sastry, {\it Proc. Natl. Acad. Sci. (US)}  {\bf 106},  3675, (2009).


\bibitem{pap:sastry-deb-stil}S. Sastry, P. G. Debenedetti and F. H. Stillinger, {\it Nature} {\bf 393}, 554 (1998).


\bibitem{pap:Sastry-pcc} S. Sastry, {\it PhysChemComm}, {\bf 3}, 79, (2000).


\bibitem{thesis:Karmakar}S. Karmakar, Ph. D. Thesis (2008).


\bibitem{roland}C. M. Roland, S. Hensel-Bielowka, M. Paluch and R. Casalini, {\it Rep. Prog. Phys.} {\bf 68}, 1405 (2005); C. M. Roland, {\it Macromolecules} {\bf 43}, 7875 (2010).



\bibitem{pap:Shi-etal}Z. Shi, P. G. Debenedetti, F. H. Stillinger and P. Ginart, {\it J. Chem. Phys.} {\bf 135}, 084513 (2011).


\end{thebibliography}
\end{document}